\documentclass[10pt,aps,prd,twocolumn,superscriptaddress,nofootinbib,floatfix,noshowpacs]{revtex4-1}

\usepackage[utf8]{inputenc}

\usepackage{diagbox}

\usepackage{mathtools}
\usepackage{amsfonts}
\usepackage{mathrsfs}
\usepackage{bbm}
\usepackage{slashed}

\usepackage{graphicx}
\usepackage{color}
\usepackage{array}
\usepackage{esint}
\usepackage{placeins}
\usepackage{booktabs}
\usepackage{makecell}
\usepackage{epstopdf}
\usepackage[caption=false]{subfig}

\usepackage{xspace}
\usepackage{siunitx}
\usepackage{hyperref}
\usepackage[nameinlink]{cleveref}
\usepackage{appendix}

\usepackage{xifthen}
\usepackage{xcolor}
\hypersetup{
	colorlinks,
	linkcolor={red!75!black},
	citecolor={blue!75!black},
	urlcolor={blue!75!black}
}

\usepackage{comment}
\usepackage{dsfont}
\usepackage{tensor}
\usepackage{tabularx}

\begin{document}


\title{Pion Valence-Quark TMD from Continuum Schwinger Function Methods and Gaussian GTMD}


\author{Minghui Ding}
\email{mhding@nju.edu.cn}
\affiliation{School of Physics, Nanjing University, Nanjing, Jiangsu 210093, China}
\affiliation{Institute for Nonperturbative Physics, Nanjing University, Nanjing, Jiangsu 210093, China}



\date{\today}

\begin{abstract}

We employ the continuum Schwinger function method to investigate the unpolarized valence-quark transverse-momentum-dependent parton distribution function (TMD) of the pion at the hadron scale. The first seventeen generalized Mellin-transverse moments, constructed from lightlike and transverse vectors, are computed and found to be well described by a factorized ansatz, in which the longitudinal component coincides with the distribution function (DF) and the transverse momentum follows a Gaussian form. The Gaussianity relation between the mean and mean-squared transverse momenta is satisfied with approximately $99\%$ accuracy in our numerical results, with the mean-squared transverse momentum equal to $0.231\,\text{GeV}^2$. Using the extracted TMD, we test the hypothesis that the quark's transverse spatial distribution also follows a Gaussian form and find that the resulting electromagnetic form factor is in good agreement with existing data. These results indicate that the intrinsic transverse-momentum and transverse-spatial distributions of valence quarks in the pion can be accurately approximated by a Gaussian ansatz, supporting its use in phenomenological analyses and experimental fits.

\end{abstract}


\maketitle



%

\section{Introduction}

The transverse structure of hadrons encodes crucial information about their three-dimensional dynamics in quantum chromodynamics (QCD). Transverse-momentum-dependent parton distribution functions (TMDs) provide direct access to this structure, extending conventional parton distribution functions (DFs) by incorporating the transverse momentum $k_T = |\mathbf{k}_T|$ of quarks and gluons~\cite{Angeles-Martinez:2015sea,Diehl:2015uka,Boussarie:2023izj}. In particular, the leading-twist unpolarized TMD, denoted here by $f_1(x,k_T^2)$, represents the probability of finding a quark with longitudinal momentum fraction $x$ and transverse momentum $k_T$. It plays a central role in the description of semi-inclusive deep inelastic scattering (SIDIS) and Drell-Yan processes~\cite{Barone:2010zz,Peng:2014hta,COMPASS:2017jbv}.  

In the perturbative region, $k_T \gg \Lambda_{\rm QCD}$, the unpolarized valence-quark TMD $f_1(x,k_T^2)$ develops a characteristic power-law tail generated by hard gluon radiation~\cite{Bacchetta:2008xw}. At leading order, it matches the collinear DFs according to
\begin{equation}
\label{eq:tmdatlargekT}
f_1(x,k_T^2)
\overset{k_T \gg \Lambda_{\rm QCD}}{\sim}
\frac{\alpha_s}{k_T^2}\,\left[ P_{qq} \otimes \mathfrak{q}(x) + P_{qg} \otimes g(x) \right],
\end{equation}
where $\mathfrak{q}(x)$ and $g(x)$ are the valence-quark and gluon collinear DFs, $P_{qq}$ and $P_{qg}$ are the leading-order Dokshitzer-Gribov-Lipatov-Altarelli-Parisi (DGLAP) splitting kernels, and $\otimes$ denotes the standard convolution in the longitudinal momentum fraction $x$. The $1/k_T^2$ dependence originates from the gluon propagator in the perturbative real-emission contribution generating transverse momentum, while the scale dependence enters through the running coupling $\alpha_s$. Equation~\eqref{eq:tmdatlargekT} demonstrates that the large-$k_T$ behavior of the TMD is perturbatively calculable, whereas the region $k_T \lesssim \Lambda_{\rm QCD}$ is intrinsically nonperturbative.  

In the nonperturbative region, the transverse structure reflects the intrinsic motion of confined quarks. Lattice QCD~\cite{Bollweg:2025iol} and model studies - including the bag model~\cite{Avakian:2010br}, the chiral quark soliton model~\cite{Wakamatsu:2009fn}, and the light-cone constituent quark model~\cite{Pasquini:2008ax} - suggest an approximate factorization between the longitudinal momentum fraction $x$ and transverse momentum $k_T$,
\begin{equation}
f_1(x,k_T^2) \overset{k_T \lesssim \Lambda_{\rm QCD}}{\sim} \mathfrak{q}(x)\, K(k_T^2),
\end{equation}
with $K(k_T^2)$ well approximated by a Gaussian profile,
\begin{equation}
K(k_T^2) = \frac{e^{-k_T^2/\langle k_T^2 \rangle}}{\pi \langle k_T^2 \rangle},
\end{equation}
where $\langle k_T^2 \rangle$ denotes the mean intrinsic transverse momentum squared. The Gaussian form is analytically simple and properly normalized. Although alternative parametrizations, such as non-Gaussian models~\cite{Radyushkin:2016hsy} or $q$-Gaussian forms~\cite{Bacchetta:2019sam}, have been explored, global analyses confirm that the Gaussian provides a robust empirical description of SIDIS and Drell-Yan data across accessible kinematics~\cite{Schweitzer:2010tt,Anselmino:2015sxa,Anselmino:2016uie,Cammarota:2020qcw,Anselmino:2020vlp}.  

Despite its phenomenological success, the Gaussian dependence lacks a derivation from QCD. Whether it emerges dynamically from nonperturbative interactions or merely represents an effective parametrization remains an open question. Addressing this issue requires a theoretically clean system governed by fundamental QCD dynamics. The pion provides such a testing ground: as the Nambu-Goldstone boson of dynamical chiral symmetry breaking and the lightest QCD bound state, it embodies both confinement and dynamical chiral symmetry breaking~\cite{Maris:1997hd,Roberts:2021nhw}. Moreover, pion-induced Drell-Yan experiments directly probe its TMDs, making it an ideal system for exploring the intrinsic $k_T$ dependence of $f_1(x,k_T^2)$~\cite{Ceccopieri:2018nop,Vladimirov:2019bfa,Bastami:2020asv,Cerutti:2022lmb,Barry:2023qqh}.  

To investigate the QCD origin of the Gaussian transverse structure, we employ the continuum Schwinger function method (CSM), a nonperturbative framework based on the Dyson-Schwinger equations of QCD~\cite{Roberts:1994dr,Maris:2003vk}. The CSM enables direct access to hadron structure through QCD's Schwinger functions and has successfully described various pion properties, including the parton distribution amplitude (DA)~\cite{Chang:2013pq}, the electromagnetic form factor~\cite{Chang:2013nia}, and the DF~\cite{Ding:2019lwe}. Since both the DA and DF encode the pion's longitudinal momentum structure, extending this framework to TMDs allows a unified treatment of longitudinal and transverse dynamics.  

The pion possesses two leading-twist TMDs: the unpolarized distribution $f_1(x,k_T^2)$ and the Boer-Mulders function $h_1^{\perp}(x,k_T^2)$~\cite{Cheng:2024gyv}. While $h_1^{\perp}(x,k_T^2)$ is time-reversal-odd, encodes spin-momentum correlations, and requires explicit gauge-link treatment~\cite{Gamberg:2009uk}, the present study focuses on the time-reversal-even distribution $f_1(x,k_T^2)$, providing a theoretically clean probe of intrinsic transverse dynamics.  

Our aim is to examine whether a Gaussian transverse-momentum dependence emerges naturally from QCD's nonperturbative dynamics. The analysis is carried out at the hadron scale, where strong quark-gluon interactions dominate, providing a foundation for future studies incorporating TMD evolution to higher scales. In this work, we focus on the nonperturbative region, $k_T \lesssim \Lambda_{\rm QCD}$, with extensions to include the asymptotic $1/k_T^2$ tail deferred to future work. This study considers only the valence-quark TMD, with sea-quark and gluon TMDs left for subsequent investigations. Comparison between CSM predictions and the empirical Gaussian ansatz clarifies whether the Gaussian form represents an emergent feature of QCD or merely a phenomenological approximation.  

The paper is organized as follows. Section~\ref{section:pdf} reviews the valence-quark DF and its Mellin moments. Section~\ref{section:tmd} introduces the valence-quark TMD and its Mellin-transverse moments, presenting the expressions for the moments used in subsequent calculations. Section~\ref{section:inputs} outlines the computational inputs for the TMD moments. Section~\ref{section:results} presents the numerical results, and Section~\ref{section:summary} concludes with a brief summary.

\section{Valence-Quark DF and its Mellin Moments}\label{section:pdf}

Owing to the close relation between the DF and TMD, we briefly summarize the valence-quark DF for completeness. Within the CSM, the pion's valence-quark DF is obtained directly from QCD's fundamental Schwinger functions, providing a self-consistent description of nonperturbative dynamics~\cite{Ding:2019lwe}. Recent studies indicate that, at the hadron scale, only dressed valence quarks and antiquarks contribute, while sea quarks and gluons are absent~\cite{Raya:2021zrz}. This behavior arises from the process-independent effective charge, which saturates in the infrared below a characteristic scale~\cite{Binosi:2016nme}. Accordingly, the pion at this scale is well described by its dressed valence degrees of freedom.  

The relevant distribution function is the valence-quark DF, $\mathfrak{q}(x)$. Under isospin symmetry ($m_u = m_d$), each valence quark (and antiquark) carries half of the pion's momentum, and the DF is symmetric under $x \leftrightarrow (1 - x)$, i.e., $\mathfrak{q}(x) = \mathfrak{q}(1 - x)$.  

At the hadron scale, the pion's valence-quark DF is defined as~\cite{Chang:2014lva,Ding:2019lwe}
\begin{align}
\label{eq:pdfdef}
\mathfrak{q}(x) = 
 N_{\mathfrak{q}}
 \int \frac{d^4 q}{(2\pi)^4}
 \,\delta\Big(\frac{n\cdot q}{n\cdot P} - x\Big)\,G(q,P)\,,
\end{align}
where
\begin{align}
G(q,P) &= \text{Tr}\Big\{
 n\cdot\partial_{q_+}
 \big[\bar{\Gamma}_\pi(q_+; -P)\, S(q_+)\big] \notag\\
 &\quad \times \Gamma_\pi(q_-; P)\, S(q_-)
 \Big\}\,,
\end{align}
and $N_{\mathfrak{q}} = -i\,\frac{N_c}{2\, n\cdot P}$. Here, $P$ is the pion momentum, $q_\pm = q \pm P/2$ are the quark and antiquark momenta, and all quantities are defined in Euclidean space. The light-like vector $n = (1,0,0,i)$ ($n^2 = 0$) projects the longitudinal momentum, $S(q)$ is the dressed-quark propagator, and $\Gamma_\pi(q;P)$ denotes the pion Bethe-Salpeter amplitude.  

The Mellin moments of $\mathfrak{q}(x)$,
\begin{align}
\label{eq:pdfmoment}
\langle x^n \rangle = \int_0^1 dx\, x^n\, \mathfrak{q}(x)\,,
\end{align}
characterize the global properties of the DF and connect directly to matrix elements of twist-two QCD operators. This correspondence provides a bridge between continuum methods~\cite{Ding:2019lwe}, lattice QCD~\cite{Zhang:2018nsy,Lin:2020ssv,Holligan:2024umc}, and phenomenological analyses~\cite{Novikov:2020snp,Barry:2021osv}. Substituting Eq.~\eqref{eq:pdfdef} into Eq.~\eqref{eq:pdfmoment} yields the CSM expression
\begin{align}
\label{eq:pdfmomentform}
\langle x^n \rangle
= N_{\mathfrak{q}}
\int \frac{d^4 q}{(2\pi)^4}
\left(\frac{n\cdot q}{n\cdot P}\right)^n G(q,P)\,.
\end{align}

The lowest moment, $\langle x^0 \rangle = \int_0^1 dx\, \mathfrak{q}(x) = 1$, indicates that, at the hadron scale, the pion's total momentum is carried by dressed valence quark and antiquark before sea-quark and gluon contributions appear through QCD evolution. The first moment, $\langle x^1 \rangle = \int_0^1 dx\, x\, \mathfrak{q}(x)$, represents the average longitudinal momentum fraction of the valence quark (or antiquark). Under isospin symmetry, $\mathfrak{q}(x) = \mathfrak{q}(1-x)$, ensuring $\langle x^1 \rangle = 1/2$ and an equal momentum share between the valence quark and antiquark.

Higher Mellin moments characterize the shape and width of $\mathfrak{q}(x)$ and can be computed directly from Eq.~\eqref{eq:pdfmomentform}. Reconstructing $\mathfrak{q}(x)$ from a finite set of moments provides a QCD-consistent description of the pion's longitudinal structure. This distribution serves as a key input for the leading-twist valence-quark unpolarized TMD, $f_1(x,k_T^2)$, ensuring that its longitudinal-momentum dependence remains consistent with the QCD collinear limit.

\section{Valence-Quark TMD and its Mellin-Transverse Moments}\label{section:tmd}

The pion's leading-twist valence-quark unpolarized TMD, $f_1(x,k_T^2)$, extends the valence-quark DF, $\mathfrak{q}(x)$, by incorporating the quark's intrinsic transverse momentum $k_T$. In the perturbative region, $k_T \gg \Lambda_{\text{QCD}}$, $f_1(x,k_T^2)$ exhibits a power-law falloff driven by gluon radiation, as shown in Eq.~\eqref{eq:tmdatlargekT}. In contrast, in the nonperturbative domain, $k_T \lesssim \Lambda_{\rm QCD}$, corresponding to the hadron scale, the QCD coupling becomes strong and requires a nonperturbative treatment. In this work, we focus on this nonperturbative region. By excluding the perturbative tail, the integral
\begin{align}
\label{eq:tmdandpdfwomu}
\int d^2 k_T\, f_1(x,k_T^2) = \mathfrak{q}(x)
\end{align}
remains finite, ensuring that $f_1(x,k_T^2)$ and $\mathfrak{q}(x)$ are defined self-consistently within the same framework.

Within the CSM, the pion's unpolarized valence-quark TMD is derived from the same QCD Schwinger functions that define $\mathfrak{q}(x)$, with an additional projection onto fixed transverse momentum:
\begin{align}
\label{eq:tmddef}
f_1(x,k_T^2) =& 
 N_{\mathfrak{q}}
 \int \!\frac{d^4 q}{(2\pi)^4}\,
 \delta\!\left(\frac{n\!\cdot\! q}{n\!\cdot\! P} - x\right)\notag\\
 &\times\delta^{(2)}\!\left(k_T - q_T\right)G(q,P)\,.
\end{align}
Here, $n$ projects the quark's longitudinal momentum, while $\delta^{(2)}(k_T-q_T)$ isolates the intrinsic transverse momentum, making the $k_T$ dependence explicit. Equation~\eqref{eq:tmddef} provides a parameter-free TMD at the hadron scale. The resulting $f_1(x,k_T^2)$ encodes the full nonperturbative transverse structure of the pion and allows a direct test of whether the empirically successful Gaussian form emerges dynamically from QCD.

Direct evaluation of Eq.~\eqref{eq:tmddef} is numerically demanding due to the delta functions fixing both longitudinal and transverse components of the quark momentum. While analytically straightforward, these constraints complicate numerical integration. A practical and widely adopted alternative is to compute the moments of the TMD, which are better suited for numerical treatment. This extends the familiar use of Mellin moments for DFs to include transverse moments that encode intrinsic momentum dependence.

A general definition of the  Mellin (longitudinal) and transverse moments is~\cite{Bastami:2018xqd,delRio:2024vvq}:
\begin{align}
\label{eq:tmdmoment}
\mathcal{M}^{n,m} \equiv \int_0^1 dx\, x^n 
  \int d^2 k_T\, k_T^{\,m}\, f_1(x,k_T^2)\,,
\end{align}
where $n$ and $m$ denote the longitudinal and transverse moment indices, respectively. The Mellin moments characterize the longitudinal momentum sharing among valence quarks, while the transverse moments describe the distribution of intrinsic transverse momentum. Here, $k_T = |\mathbf{k}_T|$ is the magnitude of the transverse momentum. By symmetry, vector-valued odd moments vanish when expressed as $\mathbf{k}_T^m$. However, since Eq.~\eqref{eq:tmdmoment} involves scalar powers $k_T^m$, the result is generally nonzero. A sufficient set of these generalized moments enables reconstruction of $f_1(x,k_T^2)$, providing a practical means to study the pion's nonperturbative transverse structure within the CSM at the hadron scale. 

Substituting Eq.~\eqref{eq:tmddef} into Eq.~\eqref{eq:tmdmoment} yields the CSM representation:
\begin{align}
\label{eq:tmdmomentform}
\mathcal{M}^{n,m} = N_{\mathfrak{q}}
   \int \frac{d^4 q}{(2\pi)^4}
   \left(\frac{n\cdot q}{n\cdot P}\right)^{n}
   (q_{Tn})^{m}\,G(q,P)\,,
\end{align}
where $q_{Tn} = |\mathbf{q}_T|$ is the magnitude of the transverse momentum, defined using the transverse basis vectors $n_{T_1}$ and $n_{T_2}$ as
\begin{align}
q_{Tn} &= \sqrt{(n_{T_1} \cdot q)^2 + (n_{T_2} \cdot q)^2}\,, \notag\\
n_{T_1} &= (0,1,0,0), \quad n_{T_2} = (0,0,1,0)\,.
\end{align}
The light-like vector $n$ projects the longitudinal component, while $n_{T_1}$ and $n_{T_2}$ project the transverse ones.

Extending the light-like vector $n$, commonly used for DFs and DAs, to include the transverse basis vectors $n_{T_1}$ and $n_{T_2}$ enables direct access to the intrinsic transverse momentum of dressed valence quarks. Although relatively new in nonperturbative studies, this approach provides a consistent framework for exploring the full three-dimensional hadron structure and is applicable to TMDs of any hadron. Systematic evaluation of $\mathcal{M}^{n,m}$ over various $(n,m)$ exposes both longitudinal and transverse features of the pion: higher longitudinal moments probe the shape and kurtosis of the $x$ distribution, while higher transverse moments reveal the width and tails of the intrinsic transverse-momentum profile. Together, they deliver a comprehensive nonperturbative description of the pion's TMD structure.

\section{Inputs for Computing $\mathcal{M}^{n,m}$}\label{section:inputs}

To evaluate $\mathcal{M}^{n,m}$ using Eq.~\eqref{eq:tmdmomentform}, one requires the dressed light-quark propagator and the pion Bethe-Salpeter amplitude, obtained self-consistently from the coupled Dyson-Schwinger and Bethe-Salpeter equations in the rainbow-ladder (RL) truncation~\cite{Qin:2011xq}. The RL truncation is the leading term in a systematic, symmetry-preserving expansion of QCD's interaction kernel~\cite{Bender:1996bb}, combining the rainbow approximation for the quark self-energy with the ladder approximation for the quark-antiquark scattering kernel. Once the effective interaction is specified, the RL scheme becomes closed and parameter-free, ensuring that the inputs to $\mathcal{M}^{n,m}$ are determined entirely by QCD dynamics.

The RL kernel is
\begin{align}
\mathcal{K}_{\alpha\beta,\gamma\delta} 
= \mathcal{G}_{\mu\nu}(k)\,[i\gamma_\mu]_{\alpha\beta}\,[i\gamma_\nu]_{\gamma\delta}\,,
\end{align}
with the effective gluon interaction modeled as
\begin{align}
\mathcal{G}_{\mu\nu}(k) = \tilde{\mathcal{G}}(k^2)\,T_{\mu\nu}(k), \quad
T_{\mu\nu}(k) = \delta_{\mu\nu} - \frac{k_\mu k_\nu}{k^2}\,.
\end{align}
The scalar part of the interaction reads~\cite{Qin:2011dd}
\begin{align}
\frac{1}{Z_2^2}\,\tilde{\mathcal{G}}(s)
= \frac{8\pi^2 D}{\omega^4}\,e^{-s/\omega^2}
+ \frac{8\pi^2 \gamma_m\,\mathcal{F}(s)}%
{\ln\big[\tau + (1 + s/\Lambda_{\text{QCD}}^2)^2\big]}\,,
\end{align}
where $s=k^2$, $\gamma_m = 4/[11 - (2/3) N_f]$ with $N_f=4$, $\Lambda_{\text{QCD}} = 0.234~\text{GeV}$, $\tau = e^2 - 1$, and $\mathcal{F}(s) = \{1 - \exp[-s/(4 m_t^2)]\}/s$ with $m_t = 0.5~\text{GeV}$. The infrared term represents the effective interaction responsible for generating a nonzero gluon mass in the infrared~\cite{Ferreira:2025anh}, while the ultraviolet term ensures the correct one-loop running.

The infrared strength and interaction width are governed by the parameters $D$ and $\omega$, constrained by $D\omega = \text{const.}$ to reproduce low-energy hadron observables. Phenomenologically successful values lie within $\omega \simeq 0.4$-$0.6~\text{GeV}$, with  
\begin{align}
D\,\omega = (0.82~\text{GeV})^3
\end{align}
providing a realistic description of pion properties. In this work, we adopt $\omega = 0.5~\text{GeV}$, corresponding to a standard interaction width.

The renormalization-group-invariant light-quark mass is  
\begin{align}
\hat{m} = 6.7~\text{MeV},
\end{align}
which corresponds to $m_{u,d}(\mu=2~\text{GeV}) \simeq 4.6~\text{MeV}$ at one loop, reproducing the empirical pion mass and decay constant. Renormalization is performed in a mass-independent momentum-subtraction scheme, fixed by the scalar Ward-Green-Takahashi identity~\cite{Chang:2008ec}.

The hadron scale is set to  
\begin{align}
\label{eq:hadronscale}
\mu_H = 0.331~\text{GeV},
\end{align}
consistent with Refs.~\cite{Cui:2021mom,Lu:2023yna} and corresponding to the inflection point of infrared saturation in the QCD effective charge~\cite{Ding:2022ows}. Below this scale, gluons no longer contribute dynamically, and the pion can be regarded as a two-body bound state of dressed valence quarks. At $\mu_H$, sea-quark and gluon degrees of freedom vanish, and the valence quark and antiquark carry the entire pion momentum.

In its fully general form, the valence-quark TMD depends on two evolution scales: the renormalization scale $\mu$ and the rapidity scale $\zeta$, i.e.,
\begin{align}
f_1(x,k_T^2) \;\rightarrow\; f_1(x,k_T^2;\mu,\zeta).
\end{align}
The presence of $\zeta$ reflects the fact that TMDs contain rapidity divergences, which are removed by a soft factor whose scale dependence is governed by the Collins-Soper equation. At large scales, the simultaneous $(\mu,\zeta)$ evolution generates the known broadening of the transverse-momentum distribution. In the present work, however, we perform the calculation at the hadron scale $\mu_H$ in Eq.~\eqref{eq:hadronscale}, where the QCD effective charge saturates in the infrared. In this regime, the soft factor becomes scale-independent and the TMD is effectively evaluated at a single natural rapidity scale, $\zeta = \mu_H^2$, so that we may write $f_1(x,k_T^2)$ without loss of generality. Rapidity evolution can be consistently applied later when evolving the TMD to higher scales for comparison with experimental data.

Working at the hadron scale thus provides a clean, parameter-free baseline in which all nonperturbative dynamics are encoded in the dressed valence quarks. There are no sea-quark or gluon distributions; the valence-quark TMD is the only nonvanishing component. Its computation requires only the dressed-quark propagator and the pion Bethe-Salpeter amplitude, both determined self-consistently within the RL truncation using the effective interaction specified above. This framework is therefore complete at $\mu_H$ and establishes a solid foundation for computing the pion unpolarized TMD, as well as for future extensions incorporating TMD evolution and large-$k_T$ behavior.

\section{Numerical results}\label{section:results}

\subsection{Valence-Quark DF: Mellin Moments and Profile}\label{subsection:pdfnum}

\begin{figure}[t]
    \centering
    \includegraphics[width=0.5\textwidth]{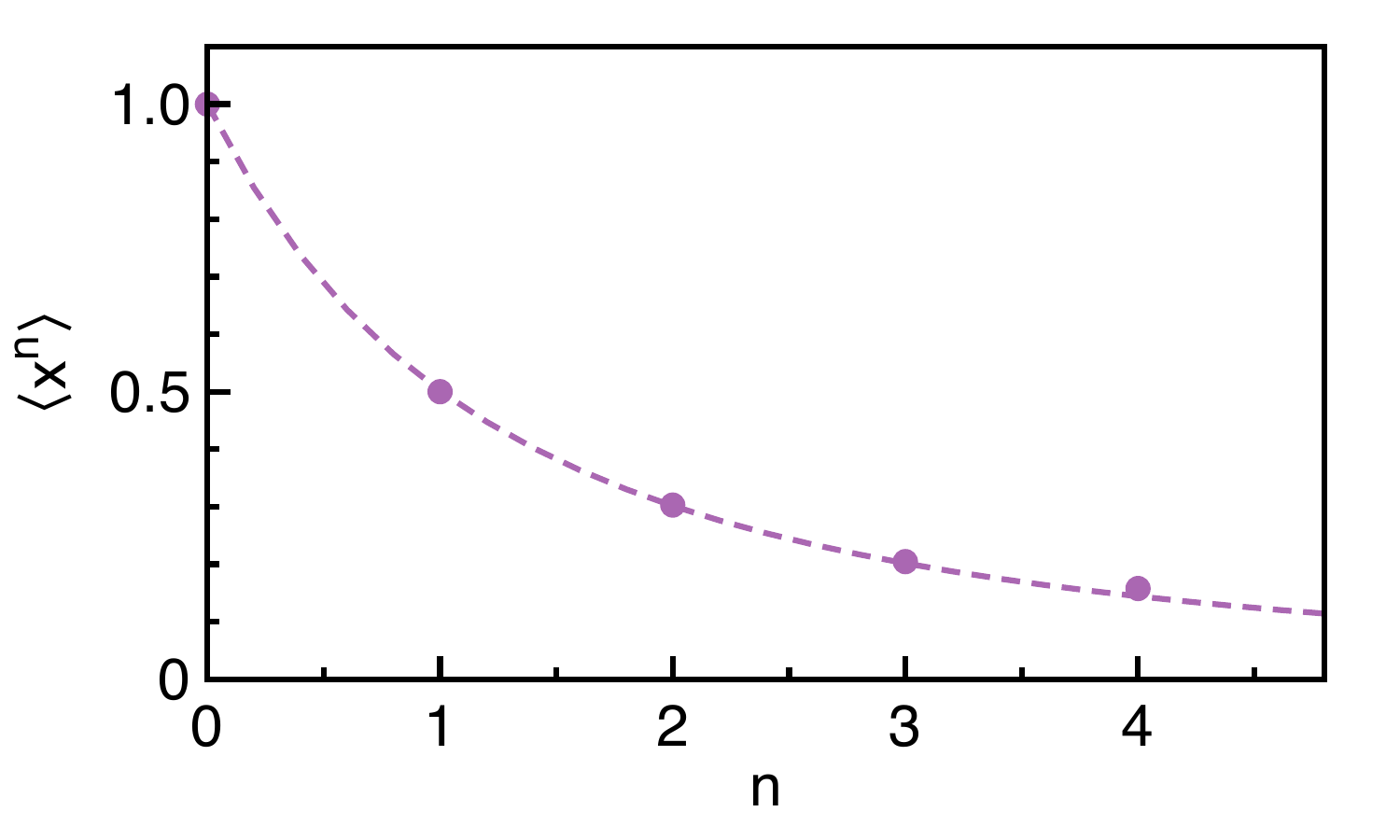}
    \caption{Pion valence-quark DF Mellin moments $\langle x^n \rangle$ ($n=0$-$4$) at the hadron scale $\mu_H$. Purple circles denote numerical results obtained from Eq.~\eqref{eq:pdfmomentform}, while the dashed curve shows the fit using Eq.~\eqref{eq:pdffunc} with $\rho_{\text{DF}}$ from Eq.~\eqref{eq:rhopdf}.}\label{fig:pdfmoments}
\end{figure}

As a baseline for the TMD analysis, we compute the Mellin moments of the pion's valence-quark DF and reconstruct the corresponding distribution.  
The lowest five moments, evaluated directly from Eq.~\eqref{eq:pdfmomentform}, are displayed in Fig.~\ref{fig:pdfmoments}.  
These moments are numerically stable and sufficient to capture the global shape of the DF at the hadron scale.  
Although higher-order moments could, in principle, be obtained using analytic continuation techniques such as the Schlessinger point method~\cite{Schlessinger:1968vsk,Schlessinger:1966zz} to refine the $x\to 1$ behavior, such refinements are not required for the present study.

The computed moments are accurately reproduced by the compact parametrization~\cite{Xu:2023bwv}
\begin{align}
\label{eq:pdffunc}
\mathfrak{q}(x)
= \frac{1}{2}\ln\!\left[1 + \frac{x^2 (1-x)^2}{\rho_{\text{DF}}^2}\right],
\end{align}
with best-fit parameter
\begin{align}
\label{eq:rhopdf}
\rho_{\text{DF}} = 0.061\,.
\end{align}
The dashed curve in Fig.~\ref{fig:pdfmoments} demonstrates the excellent agreement between this parametrization and the directly computed Mellin moments.

\begin{figure}[t]
    \centering
    \includegraphics[width=0.5\textwidth]{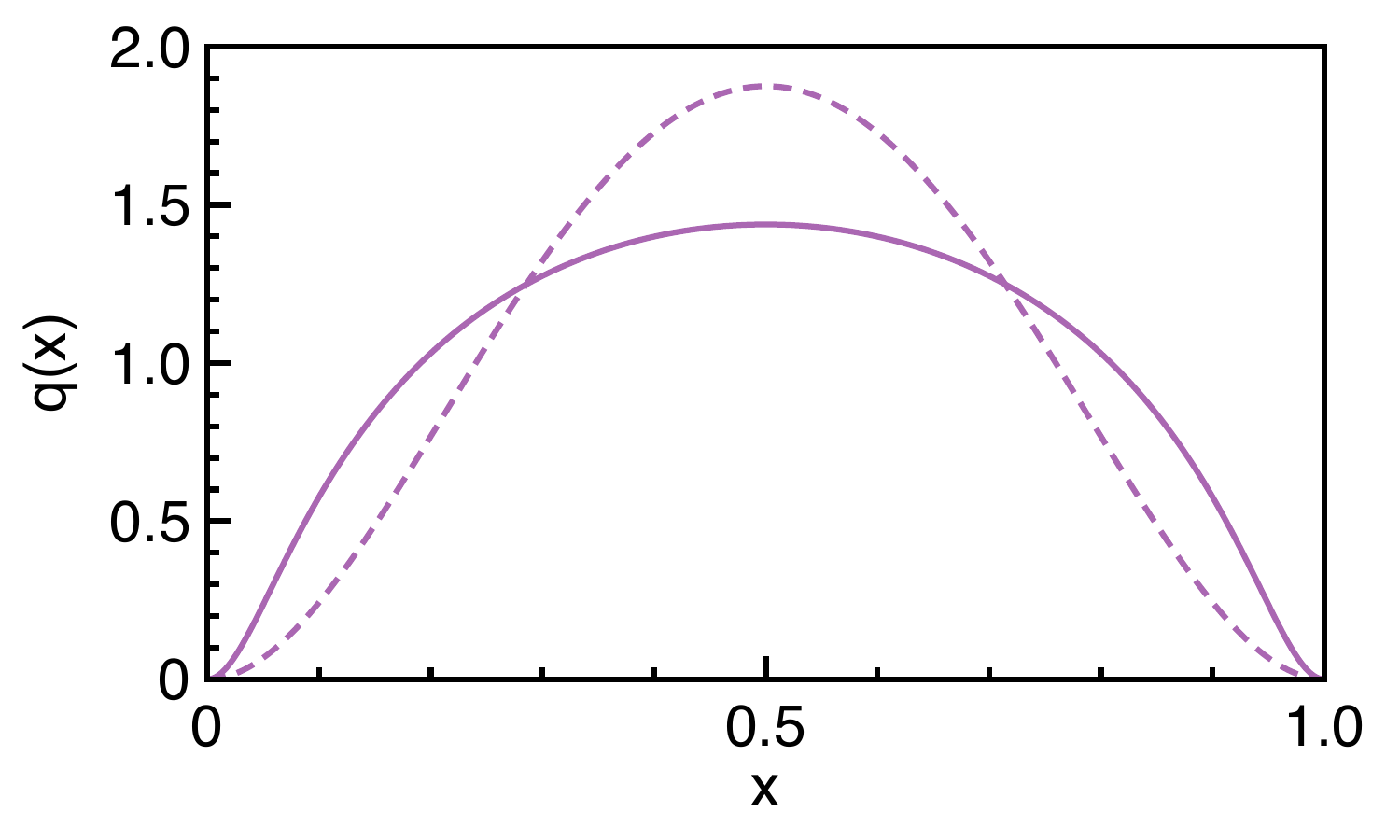}
    \caption{
        Pion valence-quark DF $\mathfrak{q}(x)$ at the hadron scale $\mu_H$. 
        The solid curve corresponds to Eq.~\eqref{eq:pdffunc} with $\rho_{\text{DF}}$ from Eq.~\eqref{eq:rhopdf}, 
        while the dashed curve shows the scale-free reference form 
        $\mathfrak{q}_{\text{sf}}(x)= 30\,x^2(1-x)^2$.
    }
    \label{fig:pdf}
\end{figure}

Equation~\eqref{eq:pdffunc} therefore provides a reliable analytic representation of the pion valence-quark DF at $\mu_H$.  
As shown in Fig.~\ref{fig:pdf}, the resulting $\mathfrak{q}(x)$ differs significantly from the scale-free form 
$\mathfrak{q}_{\text{sf}}(x)$: it is broader and more concave, reflecting the strong infrared dynamics associated with dynamical chiral symmetry breaking and the emergence of hadron mass~\cite{Roberts:2021nhw}.  
The symmetry under $x\leftrightarrow (1-x)$ confirms that, at the hadron scale, the valence quark and antiquark each carry one half of the pion's longitudinal momentum.

\subsection{Valence-Quark TMD: Mellin-Transverse Moments and Profile}

\begin{figure}[t]
    \centering
    \includegraphics[width=0.5\textwidth]{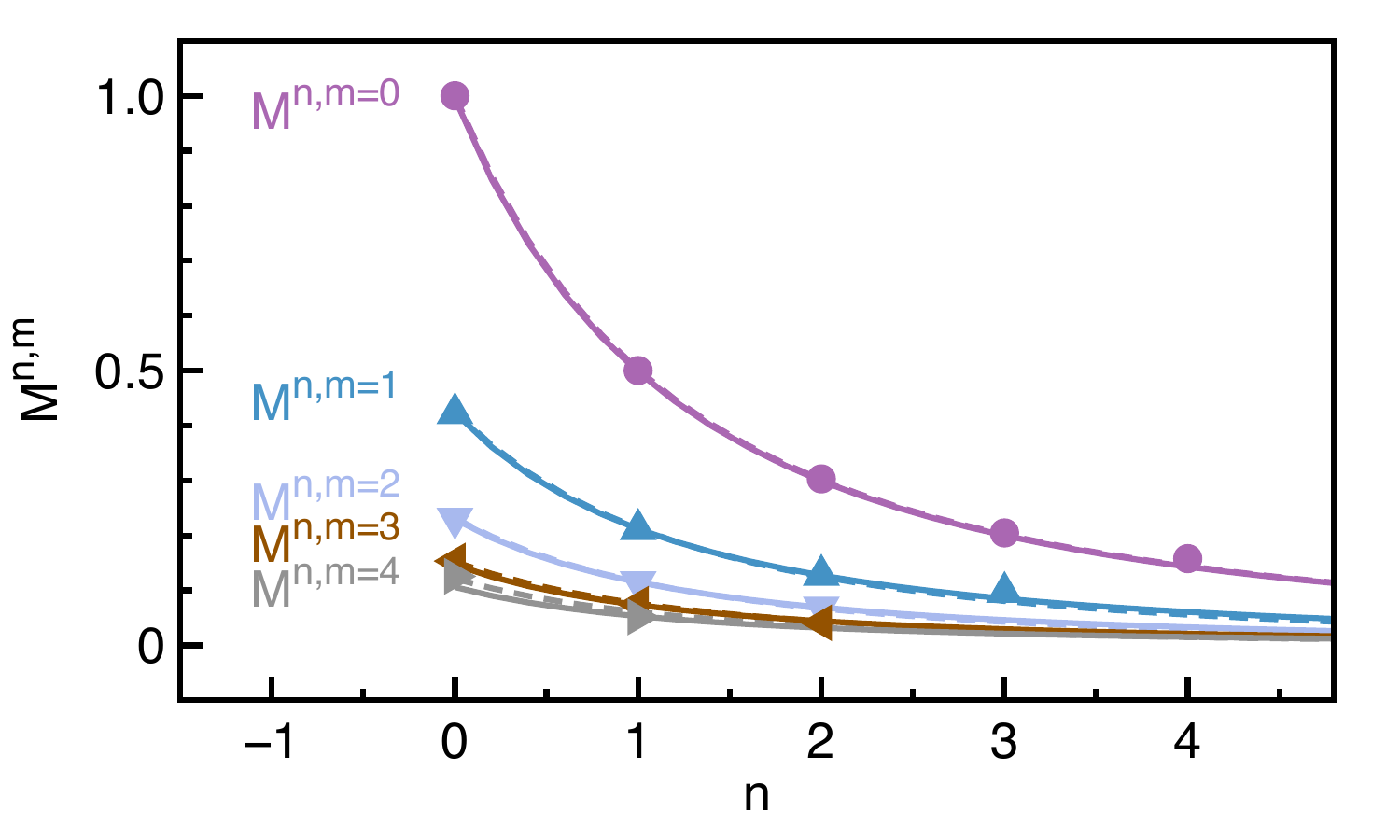}
    \caption{
        Pion generalized Mellin-transverse TMD moments $\mathcal{M}^{n,m}$ at the hadron scale $\mu_H$.  
        Symbols denote direct numerical results from Eq.~\eqref{eq:tmdmomentform}:
        circles ($m=0$), upward triangles ($m=1$), downward triangles ($m=2$), leftward triangles ($m=3$), and rightward triangles ($m=4$).  
        Dashed curves show fits using Eq.~\eqref{eq:tmdfuncmellin} with parameters $\rho_m$ from Eq.~\eqref{eq:rhotmdmellin}, and solid curves represent Gaussian fits obtained from Eq.~\eqref{eq:gaussianfunc} with $\rho_{\text{TMD}}$ in Eq.~\eqref{eq:rhotmd}.  
        Units of $\mathcal{M}^{n,m}$ are GeV$^{m}$.
    }
    \label{fig:tmdmoments}
\end{figure}

We now evaluate the generalized Mellin-transverse moments of the pion's valence-quark TMD, $\mathcal{M}^{n,m}$, defined in Eq.~\eqref{eq:tmdmomentform}.  
In practice, only a finite number of such moments can be computed with numerical stability, since higher orders require increasingly precise control of the large-$k_T$ region.  
For $m=0$, one recovers the standard DF moments, $\mathcal{M}^{n,0}=\langle x^n\rangle$, which remain reliable for $n<5$, consistent with Sec.~\ref{subsection:pdfnum}.  
As $m$ increases, the accessible range of $n$ decreases: $m=1$ permits $n<4$, $m=2$ and $m=3$ allow $n<3$, and $m=4$ is limited to $n<2$.  
This trend reflects the increasing sensitivity of higher transverse moments to the suppressed large-$k_T$ domain, where numerical uncertainties accumulate.  
Although high-$m$ moments probe the nonperturbative tail of the TMD, our primary focus is its behavior in the low-$k_T$ region.  
Altogether, seventeen moments are computed, as illustrated in Fig.~\ref{fig:tmdmoments}, which displays the systematic reduction of accessible $(n,m)$ pairs with increasing $m$.

The generalized Mellin-transverse moments may be written as
\begin{align}
\mathcal{M}^{n,m} = \int_0^1 dx\, x^n F_m(x),
\end{align}
where
\begin{align}
F_m(x) = \int d^2k_T\, k_T^m f_1(x,k_T^2).
\end{align}
For $m=0$, this reduces to $F_0(x)=\mathfrak{q}(x)$, and $\mathcal{M}^{n,0}$ recovers the DF Mellin moments.  
Since the DF is well described by Eq.~\eqref{eq:pdffunc}, it is natural to expect that the functions $F_m(x)$ for $m>0$ may follow a similar $x$ dependence.  
Extending Eq.~\eqref{eq:pdffunc}, we adopt
\begin{align}
\label{eq:tmdfuncmellin}
F_m(x) = \frac{1}{2}\ln\!\left[1+\frac{x^2(1-x)^2}{\rho_m^2}\right].
\end{align}
A fit to the computed moments yields
\begin{align}
\label{eq:rhotmdmellin}
\rho_m = \{0.061,\,0.145,\,0.227,\,0.297,\,0.340\},
\end{align}
for $m=0$-$4$.  
As shown in Fig.~\ref{fig:tmdmoments}, this parametrization accurately reproduces the numerical results for all available $(n,m)$ pairs.  
The smooth evolution of $\rho_m$ indicates that higher transverse orders preserve the qualitative $x$ dependence of the DF, with the width controlled by the increasing $k_T$ weight.

To test whether a single analytic form can describe all seventeen computed moments, we consider a factorized Ansatz in which the $x$ and $k_T$ dependences of the TMD separate.  
Motivated by the relation between the TMD and the DF, we adopt the DF functional form for the longitudinal dependence, introduce a single free parameter $\rho_{\text{TMD}}$, and assume a Gaussian transverse profile:
\begin{align}
\label{eq:gaussianfunc}
f_1(x,k_T^2) =
\frac{1}{2}\ln\!\left[1+\frac{x^2(1-x)^2}{\rho_{\text{TMD}}^2}\right]
\frac{e^{-k_T^2/\langle k_T^2\rangle}}{\pi \langle k_T^2\rangle}.
\end{align}

The transverse moments follow as
\begin{align}
\label{eq:transversemoms}
\langle k_T^m \rangle
= \mathcal{M}^{0,m}
= \int_0^1 dx \int d^2k_T\, k_T^{\,m} f_1(x,k_T^2).
\end{align}
The second moment fixes the Gaussian width.  
Our CSM calculation gives
\begin{align}
\label{eq:a}
\langle k_T^2 \rangle = \mathcal{M}^{0,2}
= 0.231~\text{GeV}^2.
\end{align}
Thus the only free parameter in Eq.~\eqref{eq:gaussianfunc} is $\rho_{\text{TMD}}$, which determines the longitudinal shape.  
We then test whether this form describes all seventeen computed moments.

A global fit of Eq.~\eqref{eq:gaussianfunc} to the computed $\mathcal{M}^{n,m}$ yields
\begin{align}
\label{eq:rhotmd}
\rho_{\text{TMD}} = \rho_{\text{DF}},
\end{align}
consistent with the collinear relation in Eq.~\eqref{eq:tmdandpdfwomu}.  
As seen in Fig.~\ref{fig:tmdmoments}, the factorized Gaussian Ansatz (solid curves) reproduces all generalized Mellin-transverse with high accuracy, indicating that it successfully captures both the longitudinal and transverse structures of the TMD obtained from the CSM.

\begin{figure}[t]
    \centering
    \includegraphics[width=0.5\textwidth]{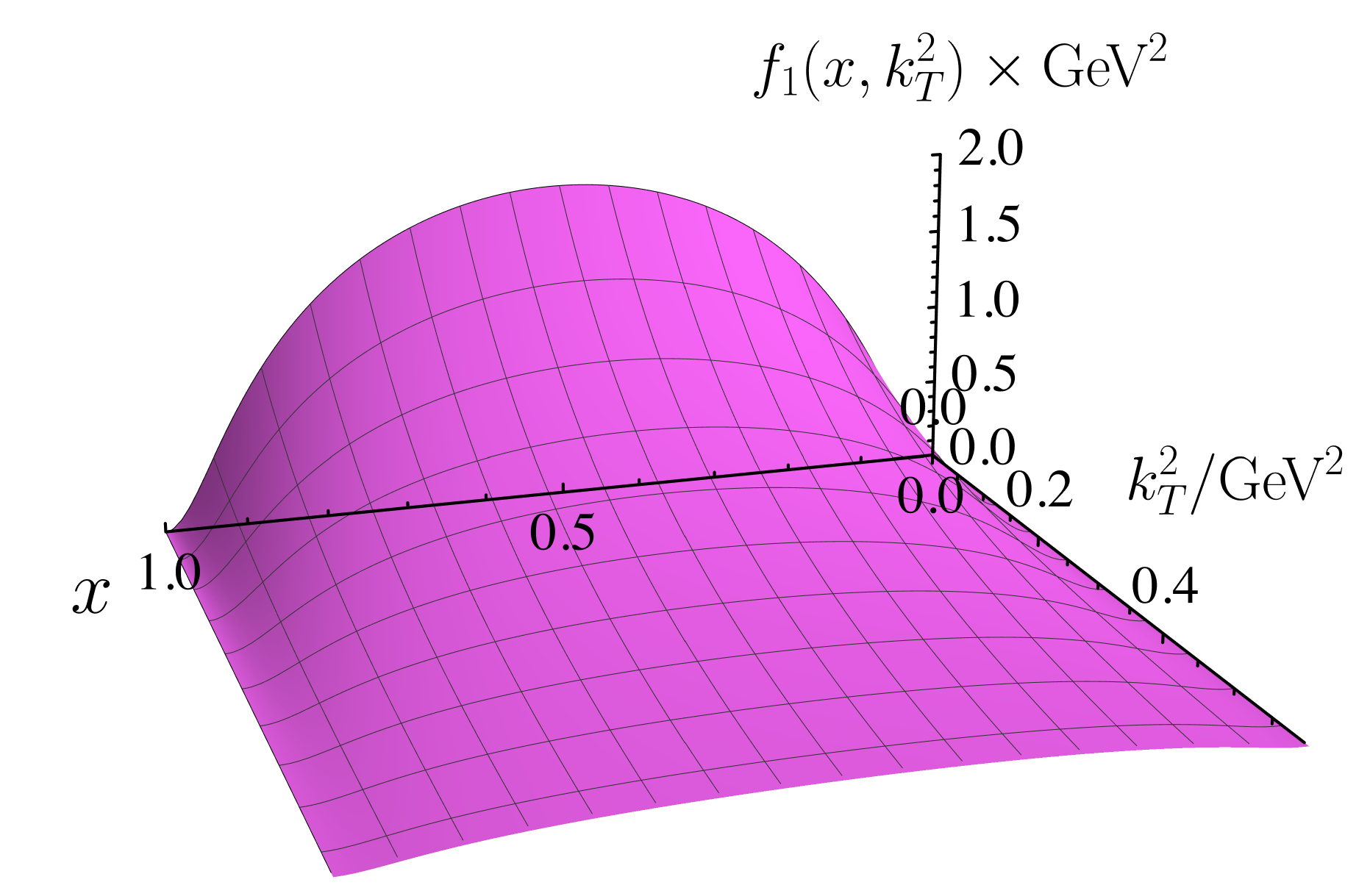}
    \caption{
        Pion valence-quark unpolarized TMD $f_1(x,k_T^2)$ at the hadron scale $\mu_H$,  
        obtained from Eq.~\eqref{eq:gaussianfunc} with $\rho_{\text{TMD}}$ in Eq.~\eqref{eq:rhotmd}.  
        The transverse dependence is Gaussian.
    }
    \label{fig:tmd}
\end{figure}

The resulting TMD is shown in Fig.~\ref{fig:tmd}.  
At $k_T^2=0$, the distribution exhibits the same broad, concave $x$ dependence as the valence-quark DF, while increasing $k_T^2$ leads to smooth Gaussian suppression.  
This demonstrates that, at the hadron scale, the intrinsic transverse momentum of the dressed valence quark in the pion is well characterized by a Gaussian profile.

\subsection{Gaussianity of the transverse-momentum distribution}

\begin{figure}[t]
    \centering
    \includegraphics[width=0.5\textwidth]{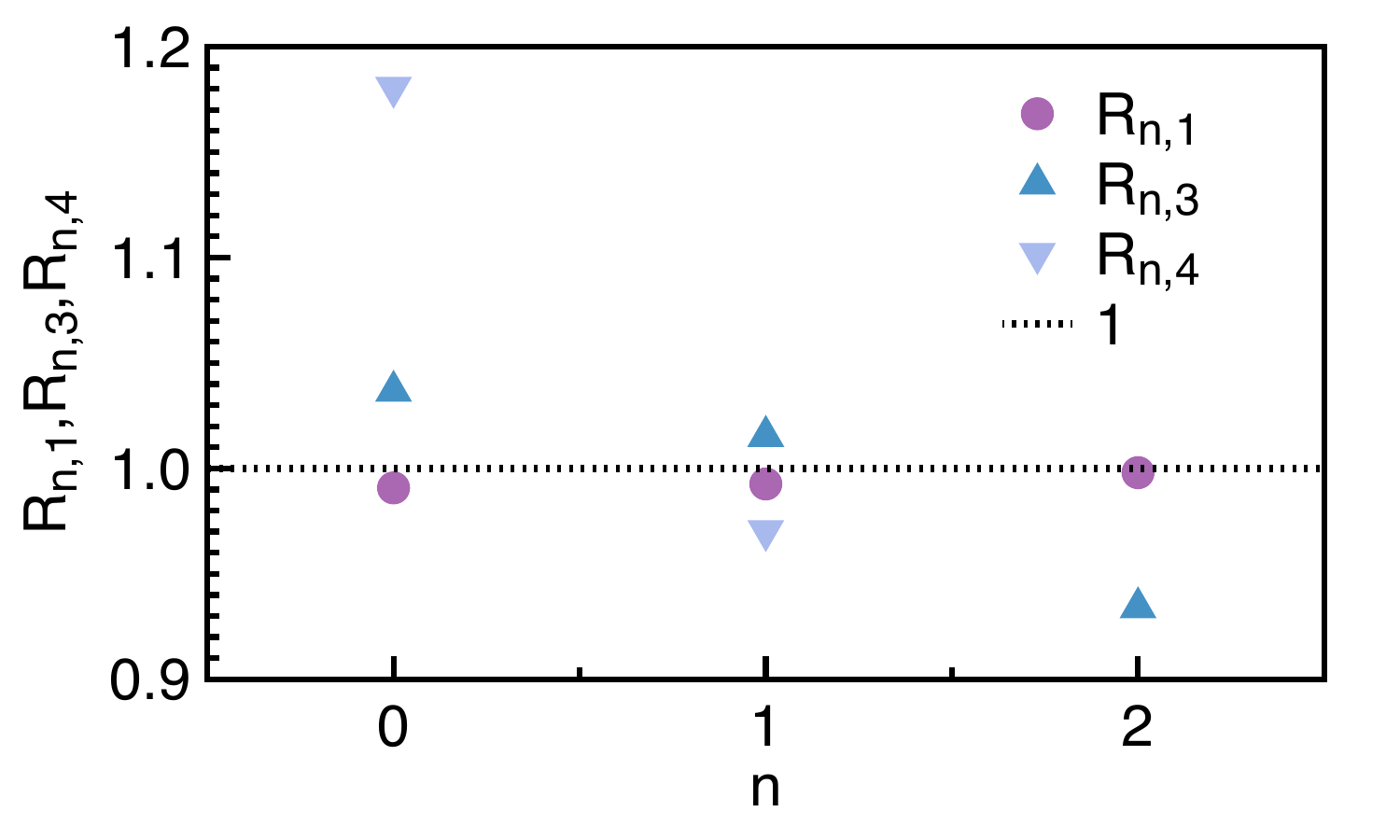}
    \caption{
        Ratios $R_{n,1}$, $R_{n,3}$, and $R_{n,4}$. 
        Symbols denote CSM results: circles ($R_{n,1}$), upward triangles ($R_{n,3}$), and downward triangles ($R_{n,4}$).
        The dotted line marks the Gaussian expectation $R_{n,m}=1$.
    }
    \label{fig:ratio}
\end{figure}

To quantify departures from a Gaussian transverse profile, we analyze the generalized Mellin-transverse moments $\mathcal{M}^{\,n,m}$ and construct dimensionless ratios.  
For each longitudinal index $n$ we define
\begin{align}
\label{eq:Rmdef}
R_{n,m}
\equiv 
\frac{1}{\Gamma\!\left(1+\frac{m}{2}\right)}
\frac{\mathcal{M}^{\,n,m}/\mathcal{M}^{\,n,0}}{\big(\mathcal{M}^{\,n,2}/\mathcal{M}^{\,n,0}\big)^{m/2}},
\end{align}
such that a purely Gaussian transverse distribution yields $R_{n,m}=1$ for all $m$. By definition $R_{n,2}\equiv1$. The case $m=1$ reduces to
\begin{align}
\label{eq:R1def}
R_{n,1}
= \frac{2}{\sqrt{\pi}}\,
\frac{\mathcal{M}^{\,n,1}/\mathcal{M}^{\,n,0}}{(\mathcal{M}^{\,n,2}/\mathcal{M}^{\,n,0})^{1/2}} .
\end{align}

For the lowest three longitudinal indices, the CSM yields
\begin{align}
R_{0,1}=0.991,\qquad
R_{1,1}=0.993,\qquad
R_{2,1}=0.998,
\end{align}
indicating deviations from unity at the level of $\lesssim1\%$.  
These values (circles in Fig.~\ref{fig:ratio}) show that the first transverse moment and the width satisfy the Gaussian relation with very high precision across all accessible $x$-moments.  
For $n=0$, where $\mathcal{M}^{0,0}=1$, $\mathcal{M}^{0,1}=\langle k_T\rangle = 0.422~\text{GeV}$ and $\mathcal{M}^{0,2}=\langle k_T^2\rangle = 0.231~\text{GeV}^2$, Eq.~\eqref{eq:R1def} gives the familiar relation
\begin{align}
R_{0,1}
= \frac{2}{\sqrt{\pi}}\,\frac{\langle k_T\rangle}{\langle k_T^2\rangle^{1/2}},
\end{align}
which leads to
\begin{align}
\langle k_T\rangle 
\simeq \left(\frac{\pi}{4}\,\langle k_T^2\rangle\right)^{1/2},
\end{align}
again confirming near-Gaussian behavior at the hadron scale.

The intrinsic transverse width $\langle k_T^2\rangle$ sets the characteristic transverse-momentum shape of the TMD.  
In the spectral quark model, evaluated at $\mu=0.313~\text{GeV}$, one finds  
$\langle k_T^2\rangle = m_\rho^2/2 \simeq 0.30~\text{GeV}^2$ with $m_\rho=0.775~\text{GeV}$~\cite{RuizArriola:2003bs}, which is larger than our result.  
Light-front constituent-quark model calculations give a smaller value,  
$\langle k_T^2\rangle = 0.102~\text{GeV}^2$~\cite{Lorce:2016ugb}, at a scale $\mu\sim0.5~\text{GeV}$.  
For comparison, phenomenological extractions of the proton unpolarized TMD yield  
$\langle k_T^2\rangle \simeq 0.25~\text{GeV}^2$~\cite{Anselmino:2005nn},  
$\langle k_T^2\rangle \simeq 0.33~\text{GeV}^2$~\cite{Collins:2005ie},  
and $\langle k_T^2\rangle = 0.38\pm0.06~\text{GeV}^2$ from HERMES data~\cite{Schweitzer:2010tt}.  
These phenomenological values correspond to scales $Q^2>2~\text{GeV}^2$, and thereby include significant evolution effects.

Our result at the hadron scale $\mu_H=0.331~\text{GeV}$,  $\langle k_T^2\rangle = 0.231~\text{GeV}^2$, lies naturally within the range suggested by QCD-based models and phenomenology.  
It should be interpreted strictly as a hadron-scale quantity: under TMD evolution, the transverse distribution broadens and its peak decreases with increasing scale~\cite{Anselmino:2012aa}.  
Since different models and extractions refer to different scales, a quantitative comparison requires careful treatment of evolution effects.

To probe higher-order structure we examine $R_{n,3}$ and $R_{n,4}$.  
For $m=3$,
\begin{align}
\label{eq:R3def}
R_{n,3}
= \frac{4}{3\sqrt{\pi}}\,
\frac{\mathcal{M}^{\,n,3}/\mathcal{M}^{\,n,0}}
     {(\mathcal{M}^{\,n,2}/\mathcal{M}^{\,n,0})^{3/2}},
\end{align}
and the CSM yields
\begin{align}
R_{0,3}=1.037,\qquad
R_{1,3}=1.015,\qquad
R_{2,3}=0.934,
\end{align}
i.e., deviations within ${\sim}7\%$ (upward triangles in Fig.~\ref{fig:ratio}).  
These values indicate very small skewness and are consistent with the expected symmetry properties of the TMD; the modest departures reflect numerical uncertainties and increasing sensitivity of higher moments to the large-$k_T$ region.

For the kurtosis-sensitive ratio ($m=4$),
\begin{align}
\label{eq:R4def}
R_{n,4}
= \frac{\mathcal{M}^{\,n,4}/\mathcal{M}^{\,n,0}}
       {2\,(\mathcal{M}^{\,n,2}/\mathcal{M}^{\,n,0})^{2}},
\end{align}
we obtain
\begin{align}
R_{0,4}=1.181,\qquad
R_{1,4}=0.970.
\end{align}
Thus $R_{0,4}$ exceeds unity by ${\sim}18\%$, while $R_{1,4}$ remains within ${\sim}3\%$ of the Gaussian value (downward triangles in Fig.~\ref{fig:ratio}).  
The enhancement of $R_{0,4}$ signals mild leptokurtic behavior: the distribution is slightly more peaked near $k_T\simeq0$ and exhibits somewhat heavier tails than a perfect Gaussian.  
Numerically, $\mathcal{M}^{0,4}_{\rm CSM}=0.126~\text{GeV}^4$ exceeds the Gaussian reference $\mathcal{M}^{0,4}_{\rm Gauss}=0.106~\text{GeV}^4$ by $0.02~\text{GeV}^4$, a small absolute shift that is amplified in the normalized ratio.

Two remarks are in order.  
First, higher transverse moments ($m\ge 4$) are increasingly sensitive to the large-$k_T$ region, where the distribution transitions from an approximately Gaussian nonperturbative core to a perturbative tail $\sim 1/k_T^2$.  
Small non-Gaussian effects in $R_{n,m}$ at large $m$ are therefore expected and do not modify the conclusion that the low-$k_T$ core is nearly Gaussian.  
Second, higher moments are numerically less stable, so percent-level deviations should be interpreted with care.  
Within the hadron-scale domain relevant here, the lower moments ($m\le 3$) provide the most reliable diagnostics and consistently support a near-Gaussian transverse structure of the pion valence-quark TMD.

Finally, the Gaussian width and the ratios $R_{n,m}$ are scale dependent.  
Under TMD evolution to higher scales, the transverse distribution broadens and the moments change accordingly~\cite{Anselmino:2012aa}.  
Thus, $\langle k_T^2\rangle$ and all $R_{n,m}$ reported here should be regarded as hadron-scale quantities that evolve with the scale.

\subsection{GTMD, GPD, and the Electromagnetic Form Factor}

The CSM moments $\mathcal{M}^{n,m}$ indicate that, at the hadron scale, the Gaussian form in Eq.~\eqref{eq:gaussianfunc} provides an excellent approximation to the valence-quark unpolarized TMD $f_1(x,k_T^2)$.  
Motivated by this observation, we model the generalized transverse-momentum-dependent distribution (GTMD) with a factorized Gaussian Ansatz,
\begin{align}
\label{eq:gtmd}
\mathcal{F}(x,k_T^2,\Delta_T^2)
 = f_1(x,k_T^2)\, e^{-B(x)\Delta_T^2},
\end{align}
where $\Delta_T$ denotes the transverse momentum transfer of the hadron and $B(x)$ controls the transverse spatial width.  
This construction ensures $\mathcal{F}(x,k^2_T,\Delta^2_T{=}0)=f_1(x,k_T^2)$ and assumes a Gaussian dependence on $\Delta_T$; possible $k_T$-$\Delta_T$ correlations are neglected as they are expected to be subleading at the hadron scale.  
The Ansatz can be tested, for example, through its implications for the pion electromagnetic form factor.

Integrating Eq.~\eqref{eq:gtmd} over $k_T$ yields the zero-skewness generalized parton distribution (GPD),
\begin{align}
H(x,0,\Delta_T^2)
 &= \int d^2k_T\, \mathcal{F}(x,k_T^2,\Delta_T^2) \notag\\
 &= \mathfrak{q}(x)\, e^{-B(x)\Delta_T^2},
\end{align}
where Eq.~\eqref{eq:tmdandpdfwomu} has been used.  
The Gaussian form of $f_1(x,k_T^2)$ ensures analytic tractability throughout.

A two-dimensional Fourier transform then gives the impact-parameter-space distribution,
\begin{align}
q(x,b_T^2)
 &= \int \frac{d^2\Delta_T}{(2\pi)^2}\,
    e^{-i b_T\cdot\Delta_T}\,
    H(x,0,\Delta_T^2) \notag\\
 &= \frac{\mathfrak{q}(x)}{4\pi B(x)}\,
    e^{-b_T^2/[4B(x)]},
\end{align}
where $b_T$ denotes the transverse position of the struck quark relative to the pion's transverse center of momentum.  
Thus, a Gaussian dependence on $\Delta_T$ leads directly to a Gaussian spatial profile in $b_T$.

The associated mean-square transverse distance is
\begin{align}
\langle b_T^2(x)\rangle
 = \frac{\int d^2 b_T\, b_T^2\, q(x,b_T^2)}%
        {\int d^2 b_T\, q(x,b_T^2)}
 = 4\,B(x).
\end{align}

A commonly used parametrization for the profile function is~\cite{Chouika:2017rzs,Raya:2021zrz,Raya:2024glv}
\begin{align}
\label{eq:bfunc}
B(x) = B_0 (1-x)^2,
\end{align}
which implies
\begin{align}
H(x,0,Q^2)
 = \mathfrak{q}(x)\, e^{-B_0 (1-x)^2 Q^2},
\end{align}
with $Q^2 \equiv \Delta_T^2$.  
Here $B_0$ controls the overall transverse size.  
This form encodes the expected correlation between longitudinal momentum and transverse localization:  
for small $x$, $(1-x)^2 \simeq 1$ and $\langle b_T^2\rangle \approx 4B_0$;  
for large $x$, $(1-x)^2 \to 0$, yielding $\langle b_T^2\rangle \to 0$.  
Fast partons are thus localized near the transverse center, whereas slow partons occupy a wider region.

\begin{figure}[t]
\centering
\includegraphics[width=0.5\textwidth]{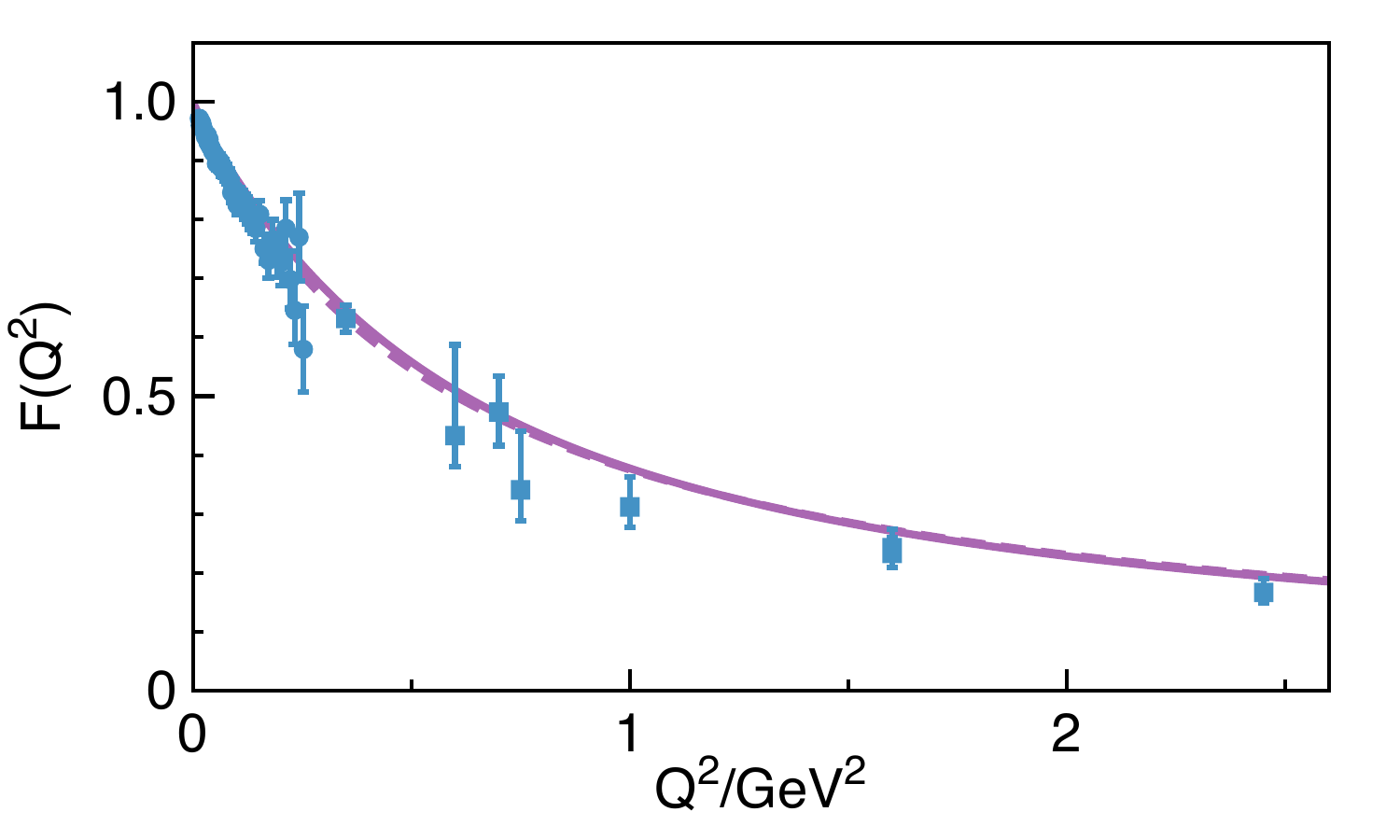}
\caption{
Pion electromagnetic form factor $F(Q^2)$.  
Solid purple curve: Eq.~\eqref{eq:formfactor} with $B_0$ from Eq.~\eqref{eq:b0value}.  
Dashed purple curve: holographic expression in Eq.~\eqref{eq:euler} with $\sqrt{\lambda}=0.548~\text{GeV}$.  
Data: circles~\cite{NA7:1986vav}, squares~\cite{JeffersonLab:2008jve}.
}
\label{fig:formfactor}
\end{figure}

The electromagnetic form factor follows from the zeroth Mellin moment of the GPD:
\begin{align}
\label{eq:formfactor}
F(Q^2)
 &= \int_0^1 dx\, H(x,0,Q^2) \notag\\
 &= \int_0^1 dx\, \mathfrak{q}(x)\,
    e^{-B_0 (1-x)^2 Q^2}.
\end{align}
Fitting to experimental data~\cite{NA7:1986vav,JeffersonLab:2008jve} fixes the single parameter,
\begin{align}
\label{eq:b0value}
B_0 = 4.699~\text{GeV}^{-2}.
\end{align}
The resulting form factor (solid curve in Fig.~\ref{fig:formfactor}) agrees well with existing measurements, supporting the Gaussian $\Delta_T$ dependence of the GTMD and the corresponding Gaussian profile in impact-parameter space.  
This demonstrates that both the transverse-momentum distribution $f_1(x,k_T^2)$ and the spatial distribution $q(x,b_T^2)$ are accurately captured by Gaussian forms at the hadron scale.

For comparison, the light-front holographic model predicts~\cite{Chang:2020kjj,Wang:2025usl}
\begin{align}
\label{eq:euler}
F(Q^2)
 = \frac{1}{2}\,
   B\!\left(1, \frac{1}{2} + \frac{Q^2}{4\lambda}\right)
 = \frac{1}{2}\int_0^1 dy\, y^{\,Q^2/(4\lambda)-1/2},
\end{align}
where $B$ denotes the Euler beta function and $\sqrt{\lambda} = 0.548~\text{GeV}=m_\rho/\sqrt{2}$.  
The holographic curve (dashed curve in Fig.~\ref{fig:formfactor}) closely tracks the Gaussian-GTMD result, reinforcing the robustness of the Gaussian description.

Beyond the electromagnetic form factor, the Gaussian-profile GTMD framework provides a unified basis for observables sensitive to both $k_T$ and $\Delta_T$.  
It can be applied to transverse-momentum-dependent form factors, spin and orbital correlations, and to processes where transverse dynamics play a central role, including SIDIS azimuthal modulations and exclusive reactions such as deeply virtual Compton scattering and deeply virtual meson production.  
It also furnishes a natural starting point for spatial imaging of hadron structure in transverse phase space.  
A detailed analysis of these extensions, including scale evolution and possible non-Gaussian behavior at large $k_T$, will be presented elsewhere.

\section{Summary}\label{section:summary}

We have investigated the pion's leading-twist valence-quark unpolarized TMD $f_1(x,k_T^2)$ at the hadron scale $\mu_H$ using the continuum Schwinger function method. In this framework, the dressed quark propagator and the pion Bethe-Salpeter amplitude are obtained self-consistently within the rainbow-ladder truncation and constitute the only numerical inputs. From these quantities we compute a set of generalized Mellin-transverse moments $\mathcal{M}^{n,m}$, which encode the combined dependence on longitudinal momentum fraction and intrinsic transverse momentum. To define these moments, a lightlike vector specifies the longitudinal direction while two orthogonal transverse vectors provide a basis for projecting the fully Poincar\'e-covariant CSM amplitudes onto well-defined generalized moments.

The seventeen computed moments $\mathcal{M}^{n,m}$ are accurately reproduced by a factorized form in which the $x$ dependence matches the valence-quark distribution function $\mathfrak{q}(x)$ and the transverse dependence is well described by a Gaussian profile with width
\begin{equation}
\langle k_T^2 \rangle = 0.231~\text{GeV}^2.
\end{equation}
Dimensionless ratios constructed from the generalized moments demonstrate that deviations from a pure Gaussian behavior are small. In particular, the relation
\begin{align}
\langle k_T\rangle \simeq \left(\frac{\pi}{4}\,\langle k_T^2\rangle\right)^{1/2},
\end{align}
is satisfied to better than $99\%$. This shows that an approximately Gaussian transverse-momentum dependence emerges dynamically from nonperturbative QCD at the hadron scale, rather than being imposed phenomenologically.

We also examined the consequences of the Gaussian transverse-momentum dependence for generalized parton distributions. Using the Gaussian behavior of the TMD extracted from the CSM moments as guidance, we employed a factorized Gaussian Ansatz for the GTMD, which leads to a Gaussian zero-skewness GPD and, via Fourier transformation, a Gaussian impact-parameter-space distribution. With a physically motivated transverse-width profile, we computed the pion electromagnetic form factor and fixed the single free parameter by comparison with experimental data. The resulting form factor agrees well with measurements and closely follows the holographic prediction, indicating that both the transverse-momentum and transverse-spatial distributions of valence quarks in the pion are well described by Gaussian forms at the hadron scale.

At the hadron scale, the valence-quark TMD is the only contribution to the unpolarized pion TMD, as dressed valence quarks constitute the active degrees of freedom in this regime. Consequently, our determination of $f_1(x,k_T^2)$ is complete at the hadron scale, with no sea-quark or gluon components. At higher scales, TMD evolution generates sea and gluon contributions, which will be explored in subsequent work.

The present study focuses on the nonperturbative region $k_T \lesssim \Lambda_{\rm QCD}$. A systematic extension requires incorporating the perturbative large-$k_T$ power-law tail and performing TMD evolution to experimental scales. These steps will enable direct comparison with pion-induced Drell-Yan measurements. Future investigations will also address the time-reversal-odd Boer-Mulders function, which necessitates explicit gauge-link dynamics. Together, these developments will advance a unified, QCD-based description of the pion's longitudinal and transverse structure.

\begin{acknowledgments}

MD gratefully acknowledges support from the Nanjing University Talent Start-up Fund. 

\end{acknowledgments}








\bibliography{bibreferences}

@article{Avakian:2010br,
    author = "Avakian, H. and Efremov, A. V. and Schweitzer, P. and Yuan, F.",
    title = "{The transverse momentum dependent distribution functions in the bag model}",
    eprint = "1001.5467",
    archivePrefix = "arXiv",
    primaryClass = "hep-ph",
    reportNumber = "JLAB-PHY-10-1154",
    doi = "10.1103/PhysRevD.81.074035",
    journal = "Phys. Rev. D",
    volume = "81",
    pages = "074035",
    year = "2010"
}

@article{Radyushkin:2016hsy,
    author = "Radyushkin, Anatoly",
    title = "{Nonperturbative Evolution of Parton Quasi-Distributions}",
    eprint = "1612.05170",
    archivePrefix = "arXiv",
    primaryClass = "hep-ph",
    reportNumber = "JLAB-THY-16-2406",
    doi = "10.1016/j.physletb.2017.02.019",
    journal = "Phys. Lett. B",
    volume = "767",
    pages = "314--320",
    year = "2017"
}

@article{Bacchetta:2008xw,
    author = "Bacchetta, Alessandro and Boer, Daniel and Diehl, Markus and Mulders, Piet J.",
    title = "{Matches and mismatches in the descriptions of semi-inclusive processes at low and high transverse momentum}",
    eprint = "0803.0227",
    archivePrefix = "arXiv",
    primaryClass = "hep-ph",
    reportNumber = "DESY-08-023",
    doi = "10.1088/1126-6708/2008/08/023",
    journal = "JHEP",
    volume = "08",
    pages = "023",
    year = "2008"
}

@article{Bacchetta:2019sam,
    author = "Bacchetta, Alessandro and Bertone, Valerio and Bissolotti, Chiara and Bozzi, Giuseppe and Delcarro, Filippo and Piacenza, Fulvio and Radici, Marco",
    title = "{Transverse-momentum-dependent parton distributions up to N$^{3}$LL from Drell-Yan data}",
    eprint = "1912.07550",
    archivePrefix = "arXiv",
    primaryClass = "hep-ph",
    reportNumber = "JLAB-THY-19-3121",
    doi = "10.1007/JHEP07(2020)117",
    journal = "JHEP",
    volume = "07",
    pages = "117",
    year = "2020"
}

@article{Cammarota:2020qcw,
    author = "Cammarota, Justin and Gamberg, Leonard and Kang, Zhong-Bo and Miller, Joshua A. and Pitonyak, Daniel and Prokudin, Alexei and Rogers, Ted C. and Sato, Nobuo",
    collaboration = "Jefferson Lab Angular Momentum",
    title = "{Origin of single transverse-spin asymmetries in high-energy collisions}",
    eprint = "2002.08384",
    archivePrefix = "arXiv",
    primaryClass = "hep-ph",
    reportNumber = "JLAB-THY-20-3151",
    doi = "10.1103/PhysRevD.102.054002",
    journal = "Phys. Rev. D",
    volume = "102",
    number = "5",
    pages = "054002",
    year = "2020"
}

@article{Anselmino:2016uie,
    author = "Anselmino, M. and Boglione, M. and D'Alesio, U. and Murgia, F. and Prokudin, A.",
    title = "{Study of the sign change of the Sivers function from STAR Collaboration W/Z production data}",
    eprint = "1612.06413",
    archivePrefix = "arXiv",
    primaryClass = "hep-ph",
    reportNumber = "JLAB-THY-16-2404",
    doi = "10.1007/JHEP04(2017)046",
    journal = "JHEP",
    volume = "04",
    pages = "046",
    year = "2017"
}

@article{Anselmino:2015sxa,
    author = "Anselmino, M. and Boglione, M. and D'Alesio, U. and Gonzalez Hernandez, J. O. and Melis, S. and Murgia, F. and Prokudin, A.",
    title = "{Collins functions for pions from SIDIS and new $e^+e^-$ data: a first glance at their transverse momentum dependence}",
    eprint = "1510.05389",
    archivePrefix = "arXiv",
    primaryClass = "hep-ph",
    doi = "10.1103/PhysRevD.92.114023",
    journal = "Phys. Rev. D",
    volume = "92",
    number = "11",
    pages = "114023",
    year = "2015"
}

@article{Schweitzer:2010tt,
    author = "Schweitzer, P. and Teckentrup, T. and Metz, A.",
    title = "{Intrinsic transverse parton momenta in deeply inelastic reactions}",
    eprint = "1003.2190",
    archivePrefix = "arXiv",
    primaryClass = "hep-ph",
    doi = "10.1103/PhysRevD.81.094019",
    journal = "Phys. Rev. D",
    volume = "81",
    pages = "094019",
    year = "2010"
}

@article{Bastami:2018xqd,
    author = "Bastami, S. and others",
    title = "{Semi-Inclusive Deep Inelastic Scattering in Wandzura-Wilczek-type approximation}",
    eprint = "1807.10606",
    archivePrefix = "arXiv",
    primaryClass = "hep-ph",
    reportNumber = "JLAB-THY-18-2775",
    doi = "10.1007/JHEP06(2019)007",
    journal = "JHEP",
    volume = "06",
    pages = "007",
    year = "2019"
}

@article{Anselmino:2020vlp,
    author = "Anselmino, Mauro and Mukherjee, Asmita and Vossen, Anselm",
    title = "{Transverse spin effects in hard semi-inclusive collisions}",
    eprint = "2001.05415",
    archivePrefix = "arXiv",
    primaryClass = "hep-ph",
    reportNumber = "JLAB-PHY-20-3235",
    doi = "10.1016/j.ppnp.2020.103806",
    journal = "Prog. Part. Nucl. Phys.",
    volume = "114",
    pages = "103806",
    year = "2020"
}

@article{Wakamatsu:2009fn,
    author = "Wakamatsu, M.",
    title = "{Transverse momentum distributions of quarks in the nucleon from the Chiral Quark Soliton Model}",
    eprint = "0903.1886",
    archivePrefix = "arXiv",
    primaryClass = "hep-ph",
    reportNumber = "OU-HET-625",
    doi = "10.1103/PhysRevD.79.094028",
    journal = "Phys. Rev. D",
    volume = "79",
    pages = "094028",
    year = "2009"
}

@article{Pasquini:2008ax,
    author = "Pasquini, B. and Cazzaniga, S. and Boffi, S.",
    title = "{Transverse momentum dependent parton distributions in a light-cone quark model}",
    eprint = "0806.2298",
    archivePrefix = "arXiv",
    primaryClass = "hep-ph",
    doi = "10.1103/PhysRevD.78.034025",
    journal = "Phys. Rev. D",
    volume = "78",
    pages = "034025",
    year = "2008"
}

@article{delRio:2024vvq,
    author = "del Rio, Oscar and Prokudin, Alexei and Scimemi, Ignazio and Vladimirov, Alexey",
    title = "{Transverse momentum moments}",
    eprint = "2402.01836",
    archivePrefix = "arXiv",
    primaryClass = "hep-ph",
    reportNumber = "IPARCOS-UCM-2024-007, JLAB-THY-24-3989",
    doi = "10.1103/PhysRevD.110.016003",
    journal = "Phys. Rev. D",
    volume = "110",
    number = "1",
    pages = "016003",
    year = "2024"
}

@article{Qin:2011dd,
    author = "Qin, Si-xue and Chang, Lei and Liu, Yu-xin and Roberts, Craig D. and Wilson, David J.",
    title = "{Interaction model for the gap equation}",
    eprint = "1108.0603",
    archivePrefix = "arXiv",
    primaryClass = "nucl-th",
    doi = "10.1103/PhysRevC.84.042202",
    journal = "Phys. Rev. C",
    volume = "84",
    pages = "042202",
    year = "2011"
}

@article{Cui:2021mom,
    author = "Cui, Z. -F. and Ding, Minghui and Morgado, J. M. and Raya, K. and Binosi, D. and Chang, L. and Papavassiliou, J. and Roberts, C. D. and Rodr{\'\i}guez-Quintero, J. and Schmidt, S. M.",
    title = "{Concerning pion parton distributions}",
    eprint = "2112.09210",
    archivePrefix = "arXiv",
    primaryClass = "hep-ph",
    reportNumber = "NJU-INP 053/21",
    doi = "10.1140/epja/s10050-021-00658-7",
    journal = "Eur. Phys. J. A",
    volume = "58",
    number = "1",
    pages = "10",
    year = "2022"
}

@article{Lu:2023yna,
    author = "Lu, Ya and Xu, Yin-Zhen and Raya, Kh{\'e}pani and Roberts, Craig D. and Rodr{\'\i}guez-Quintero, Jos{\'e}",
    title = "{Pion distribution functions from low-order Mellin moments}",
    eprint = "2311.08565",
    archivePrefix = "arXiv",
    primaryClass = "hep-ph",
    reportNumber = "NJU-INP 080/23",
    doi = "10.1016/j.physletb.2024.138534",
    journal = "Phys. Lett. B",
    volume = "850",
    pages = "138534",
    year = "2024"
}

@article{Roberts:2021nhw,
    author = "Roberts, Craig D. and Richards, David G. and Horn, Tanja and Chang, Lei",
    title = "{Insights into the emergence of mass from studies of pion and kaon structure}",
    eprint = "2102.01765",
    archivePrefix = "arXiv",
    primaryClass = "hep-ph",
    reportNumber = "NJU-INP 034/21",
    doi = "10.1016/j.ppnp.2021.103883",
    journal = "Prog. Part. Nucl. Phys.",
    volume = "120",
    pages = "103883",
    year = "2021"
}

@article{RuizArriola:2003bs,
    author = "Ruiz Arriola, Enrique and Broniowski, Wojciech",
    title = "{Spectral quark model and low-energy hadron phenomenology}",
    eprint = "hep-ph/0301202",
    archivePrefix = "arXiv",
    doi = "10.1103/PhysRevD.67.074021",
    journal = "Phys. Rev. D",
    volume = "67",
    pages = "074021",
    year = "2003"
}

@article{Anselmino:2012aa,
    author = "Anselmino, M. and Boglione, M. and Melis, S.",
    title = "{A Strategy towards the extraction of the Sivers function with TMD evolution}",
    eprint = "1204.1239",
    archivePrefix = "arXiv",
    primaryClass = "hep-ph",
    doi = "10.1103/PhysRevD.86.014028",
    journal = "Phys. Rev. D",
    volume = "86",
    pages = "014028",
    year = "2012"
}

@article{JeffersonLab:2008jve,
    author = "Huber, G. M. and others",
    collaboration = "Jefferson Lab",
    title = "{Charged pion form-factor between Q**2 = 0.60-GeV**2 and 2.45-GeV**2. II. Determination of, and results for, the pion form-factor}",
    eprint = "0809.3052",
    archivePrefix = "arXiv",
    primaryClass = "nucl-ex",
    reportNumber = "JLAB-PHY-08-864",
    doi = "10.1103/PhysRevC.78.045203",
    journal = "Phys. Rev. C",
    volume = "78",
    pages = "045203",
    year = "2008"
}

@article{NA7:1986vav,
    author = "Amendolia, S. R. and others",
    editor = "Loken, S. C.",
    collaboration = "NA7",
    title = "{A Measurement of the Space - Like Pion Electromagnetic Form-Factor}",
    reportNumber = "CERN-EP-86-34",
    doi = "10.1016/0550-3213(86)90437-2",
    journal = "Nucl. Phys. B",
    volume = "277",
    pages = "168",
    year = "1986"
}

@article{Chang:2008ec,
    author = "Chang, Lei and Liu, Yu-xin and Roberts, Craig D. and Shi, Yuan-mei and Sun, Wei-min and Zong, Hong-shi",
    title = "{Chiral susceptibility and the scalar Ward identity}",
    eprint = "0812.2956",
    archivePrefix = "arXiv",
    primaryClass = "nucl-th",
    reportNumber = "ANL-PHY-12242-TH-2008",
    doi = "10.1103/PhysRevC.79.035209",
    journal = "Phys. Rev. C",
    volume = "79",
    pages = "035209",
    year = "2009"
}

@article{Chouika:2017rzs,
    author = "Chouika, N. and Mezrag, C. and Moutarde, H. and Rodr{\'\i}guez-Quintero, J.",
    title = "{A Nakanishi-based model illustrating the covariant extension of the pion GPD overlap representation and its ambiguities}",
    eprint = "1711.11548",
    archivePrefix = "arXiv",
    primaryClass = "hep-ph",
    doi = "10.1016/j.physletb.2018.02.070",
    journal = "Phys. Lett. B",
    volume = "780",
    pages = "287--293",
    year = "2018"
}

@article{Boussarie:2023izj,
    author = "Boussarie, Renaud and others",
    title = "{TMD Handbook}",
    eprint = "2304.03302",
    archivePrefix = "arXiv",
    primaryClass = "hep-ph",
    reportNumber = "JLAB-THY-23-3780, LA-UR-21-20798, MIT-CTP/5386",
    month = "4",
    year = "2023"
}

@article{Diehl:2015uka,
    author = "Diehl, Markus",
    title = "{Introduction to GPDs and TMDs}",
    eprint = "1512.01328",
    archivePrefix = "arXiv",
    primaryClass = "hep-ph",
    reportNumber = "DESY-15-234",
    doi = "10.1140/epja/i2016-16149-3",
    journal = "Eur. Phys. J. A",
    volume = "52",
    number = "6",
    pages = "149",
    year = "2016"
}

@article{Barone:2010zz,
    author = "Barone, Vincenzo and Bradamante, Franco and Martin, Anna",
    title = "{Transverse-spin and transverse-momentum effects in high-energy processes}",
    eprint = "1011.0909",
    archivePrefix = "arXiv",
    primaryClass = "hep-ph",
    doi = "10.1016/j.ppnp.2010.07.003",
    journal = "Prog. Part. Nucl. Phys.",
    volume = "65",
    pages = "267--333",
    year = "2010"
}

@article{Angeles-Martinez:2015sea,
    author = "Angeles-Martinez, R. and others",
    title = "{Transverse Momentum Dependent (TMD) parton distribution functions: status and prospects}",
    eprint = "1507.05267",
    archivePrefix = "arXiv",
    primaryClass = "hep-ph",
    reportNumber = "DESY-15-111, NIKHEF-2015-023, RAL-P-2015-006, JLAB-THY-15-2020",
    doi = "10.5506/APhysPolB.46.2501",
    journal = "Acta Phys. Polon. B",
    volume = "46",
    number = "12",
    pages = "2501--2534",
    year = "2015"
}

@article{COMPASS:2017jbv,
    author = "Aghasyan, M. and others",
    collaboration = "COMPASS",
    title = "{First measurement of transverse-spin-dependent azimuthal asymmetries in the Drell-Yan process}",
    eprint = "1704.00488",
    archivePrefix = "arXiv",
    primaryClass = "hep-ex",
    reportNumber = "CERN-EP-2017-059",
    doi = "10.1103/PhysRevLett.119.112002",
    journal = "Phys. Rev. Lett.",
    volume = "119",
    number = "11",
    pages = "112002",
    year = "2017"
}

@article{Peng:2014hta,
    author = "Peng, Jen-Chieh and Qiu, Jian-Wei",
    title = "{Novel phenomenology of parton distributions from the Drell{\textendash}Yan process}",
    eprint = "1401.0934",
    archivePrefix = "arXiv",
    primaryClass = "hep-ph",
    doi = "10.1016/j.ppnp.2014.01.005",
    journal = "Prog. Part. Nucl. Phys.",
    volume = "76",
    pages = "43--75",
    year = "2014"
}

@article{Vladimirov:2019bfa,
    author = "Vladimirov, Alexey",
    title = "{Pion-induced Drell-Yan processes within TMD factorization}",
    eprint = "1907.10356",
    archivePrefix = "arXiv",
    primaryClass = "hep-ph",
    doi = "10.1007/JHEP10(2019)090",
    journal = "JHEP",
    volume = "10",
    pages = "090",
    year = "2019"
}

@article{Bastami:2020asv,
    author = "Bastami, S. and Gamberg, L. and Parsamyan, B. and Pasquini, B. and Prokudin, A. and Schweitzer, P.",
    title = "{The Drell-Yan process with pions and polarized nucleons}",
    eprint = "2005.14322",
    archivePrefix = "arXiv",
    primaryClass = "hep-ph",
    reportNumber = "JLAB-THY-20-3199",
    doi = "10.1007/JHEP02(2021)166",
    journal = "JHEP",
    volume = "02",
    pages = "166",
    year = "2021"
}

@article{Cerutti:2022lmb,
    author = "Cerutti, Matteo and Rossi, Lorenzo and Venturini, Simone and Bacchetta, Alessandro and Bertone, Valerio and Bissolotti, Chiara and Radici, Marco",
    collaboration = "MAP (Multi-dimensional Analyses of Partonic distributions)",
    title = "{Extraction of pion transverse momentum distributions from Drell-Yan data}",
    eprint = "2210.01733",
    archivePrefix = "arXiv",
    primaryClass = "hep-ph",
    doi = "10.1103/PhysRevD.107.014014",
    journal = "Phys. Rev. D",
    volume = "107",
    number = "1",
    pages = "014014",
    year = "2023"
}

@article{Ceccopieri:2018nop,
    author = "Ceccopieri, Federico Alberto and Courtoy, Aurore and Noguera, Santiago and Scopetta, Sergio",
    title = "{Pion nucleus Drell{\textendash}Yan process and parton transverse momentum in the pion}",
    eprint = "1801.07682",
    archivePrefix = "arXiv",
    primaryClass = "hep-ph",
    doi = "10.1140/epjc/s10052-018-6115-3",
    journal = "Eur. Phys. J. C",
    volume = "78",
    number = "8",
    pages = "644",
    year = "2018"
}

@article{Barry:2023qqh,
    author = "Barry, P. C. and Gamberg, L. and Melnitchouk, W. and Moffat, E. and Pitonyak, D. and Prokudin, A. and Sato, N.",
    collaboration = "Jefferson Lab Angular Momentum (JAM)",
    title = "{Tomography of pions and protons via transverse momentum dependent distributions}",
    eprint = "2302.01192",
    archivePrefix = "arXiv",
    primaryClass = "hep-ph",
    reportNumber = "JLAB-THY-23-3749, ADP-23-03/T1212",
    doi = "10.1103/PhysRevD.108.L091504",
    journal = "Phys. Rev. D",
    volume = "108",
    number = "9",
    pages = "L091504",
    year = "2023"
}

@article{Maris:1997hd,
    author = "Maris, Pieter and Roberts, Craig D. and Tandy, Peter C.",
    title = "{Pion mass and decay constant}",
    eprint = "nucl-th/9707003",
    archivePrefix = "arXiv",
    reportNumber = "ANL-PHY-8753-TH-97, KSUCNR-103-97",
    doi = "10.1016/S0370-2693(97)01535-9",
    journal = "Phys. Lett. B",
    volume = "420",
    pages = "267--273",
    year = "1998"
}

@article{Roberts:1994dr,
    author = "Roberts, Craig D. and Williams, Anthony G.",
    title = "{Dyson-Schwinger equations and their application to hadronic physics}",
    eprint = "hep-ph/9403224",
    archivePrefix = "arXiv",
    reportNumber = "ADP-93-225-T-142, ANL-PHY-7668-TH-93",
    doi = "10.1016/0146-6410(94)90049-3",
    journal = "Prog. Part. Nucl. Phys.",
    volume = "33",
    pages = "477--575",
    year = "1994"
}

@article{Maris:2003vk,
    author = "Maris, Pieter and Roberts, Craig D.",
    title = "{Dyson-Schwinger equations: A Tool for hadron physics}",
    eprint = "nucl-th/0301049",
    archivePrefix = "arXiv",
    reportNumber = "ANL-PHY-10465-TH-2002",
    doi = "10.1142/S0218301303001326",
    journal = "Int. J. Mod. Phys. E",
    volume = "12",
    pages = "297--365",
    year = "2003"
}

@article{Chang:2013pq,
    author = "Chang, Lei and Cloet, I. C. and Cobos-Martinez, J. J. and Roberts, C. D. and Schmidt, S. M. and Tandy, P. C.",
    title = "{Imaging dynamical chiral symmetry breaking: pion wave function on the light front}",
    eprint = "1301.0324",
    archivePrefix = "arXiv",
    primaryClass = "nucl-th",
    doi = "10.1103/PhysRevLett.110.132001",
    journal = "Phys. Rev. Lett.",
    volume = "110",
    number = "13",
    pages = "132001",
    year = "2013"
}

@article{Chang:2013nia,
    author = {Chang, L. and Clo{\"e}t, I. C. and Roberts, C. D. and Schmidt, S. M. and Tandy, P. C.},
    title = "{Pion electromagnetic form factor at spacelike momenta}",
    eprint = "1307.0026",
    archivePrefix = "arXiv",
    primaryClass = "nucl-th",
    doi = "10.1103/PhysRevLett.111.141802",
    journal = "Phys. Rev. Lett.",
    volume = "111",
    number = "14",
    pages = "141802",
    year = "2013"
}

@article{Ding:2019lwe,
    author = "Ding, Minghui and Raya, Kh{\'e}pani and Binosi, Daniele and Chang, Lei and Roberts, Craig D and Schmidt, Sebastian M.",
    title = "{Symmetry, symmetry breaking, and pion parton distributions}",
    eprint = "1905.05208",
    archivePrefix = "arXiv",
    primaryClass = "nucl-th",
    reportNumber = "NJU-INP 003/19",
    doi = "10.1103/PhysRevD.101.054014",
    journal = "Phys. Rev. D",
    volume = "101",
    number = "5",
    pages = "054014",
    year = "2020"
}

@article{Cheng:2024gyv,
    author = "Cheng, Dan-Dan and Cui, Zhu-Fang and Ding, Minghui and Roberts, Craig D. and Schmidt, Sebastian M.",
    title = "{Pion Boer{\textendash}Mulders function using a contact interaction}",
    eprint = "2409.11568",
    archivePrefix = "arXiv",
    primaryClass = "hep-ph",
    reportNumber = "NJU-INP 090/24, NJU-INP 090/23",
    doi = "10.1140/epjc/s10052-025-13782-1",
    journal = "Eur. Phys. J. C",
    volume = "85",
    number = "1",
    pages = "115",
    year = "2025"
}

@article{Gamberg:2009uk,
    author = "Gamberg, Leonard and Schlegel, Marc",
    title = "{Final state interactions and the transverse structure of the pion using non-perturbative eikonal methods}",
    eprint = "0911.1964",
    archivePrefix = "arXiv",
    primaryClass = "hep-ph",
    reportNumber = "INT-PUB-09-056, JLAB-THY-09-1104",
    doi = "10.1016/j.physletb.2009.12.067",
    journal = "Phys. Lett. B",
    volume = "685",
    pages = "95--103",
    year = "2010"
}

@article{Chang:2014lva,
    author = "Chang, Lei and Mezrag, C{\'e}dric and Moutarde, Herv{\'e} and Roberts, Craig D. and Rodr{\'\i}guez-Quintero, Jose and Tandy, Peter C.",
    title = "{Basic features of the pion valence-quark distribution function}",
    eprint = "1406.5450",
    archivePrefix = "arXiv",
    primaryClass = "nucl-th",
    reportNumber = "ADP-14-20-T878",
    doi = "10.1016/j.physletb.2014.08.009",
    journal = "Phys. Lett. B",
    volume = "737",
    pages = "23--29",
    year = "2014"
}

@article{Zhang:2018nsy,
    author = {Zhang, Jian-Hui and Chen, Jiunn-Wei and Jin, Luchang and Lin, Huey-Wen and Sch{\"a}fer, Andreas and Zhao, Yong},
    title = "{First direct lattice-QCD calculation of the $x$-dependence of the pion parton distribution function}",
    eprint = "1804.01483",
    archivePrefix = "arXiv",
    primaryClass = "hep-lat",
    doi = "10.1103/PhysRevD.100.034505",
    journal = "Phys. Rev. D",
    volume = "100",
    number = "3",
    pages = "034505",
    year = "2019"
}

@article{Holligan:2024umc,
    author = "Holligan, Jack and Lin, Huey-Wen",
    title = "{Pion valence quark distribution at physical pion mass of N $_{f}$ = 2 + 1 + 1 lattice QCD}",
    eprint = "2404.14525",
    archivePrefix = "arXiv",
    primaryClass = "hep-lat",
    reportNumber = "MSUHEP-23-032",
    doi = "10.1088/1361-6471/ad3162",
    journal = "J. Phys. G",
    volume = "51",
    number = "6",
    pages = "065101",
    year = "2024"
}

@article{Lin:2020ssv,
    author = "Lin, Huey-Wen and Chen, Jiunn-Wei and Fan, Zhouyou and Zhang, Jian-Hui and Zhang, Rui",
    title = "{Valence-Quark Distribution of the Kaon and Pion from Lattice QCD}",
    eprint = "2003.14128",
    archivePrefix = "arXiv",
    primaryClass = "hep-lat",
    reportNumber = "MSUHEP-20-006",
    doi = "10.1103/PhysRevD.103.014516",
    journal = "Phys. Rev. D",
    volume = "103",
    number = "1",
    pages = "014516",
    year = "2021"
}

@article{Barry:2021osv,
    author = "Barry, P. C. and Ji, Chueng-Ryong and Sato, N. and Melnitchouk, W.",
    collaboration = "Jefferson Lab Angular Momentum (JAM)",
    title = "{Global QCD Analysis of Pion Parton Distributions with Threshold Resummation}",
    eprint = "2108.05822",
    archivePrefix = "arXiv",
    primaryClass = "hep-ph",
    reportNumber = "JLAB-THY-21-3482",
    doi = "10.1103/PhysRevLett.127.232001",
    journal = "Phys. Rev. Lett.",
    volume = "127",
    number = "23",
    pages = "232001",
    year = "2021"
}

@article{Novikov:2020snp,
    author = "Novikov, Ivan and others",
    title = "{Parton Distribution Functions of the Charged Pion Within The xFitter Framework}",
    eprint = "2002.02902",
    archivePrefix = "arXiv",
    primaryClass = "hep-ph",
    reportNumber = "DESY-20-013, DESY 20-013",
    doi = "10.1103/PhysRevD.102.014040",
    journal = "Phys. Rev. D",
    volume = "102",
    number = "1",
    pages = "014040",
    year = "2020"
}

@article{Raya:2021zrz,
    author = "Raya, Khepani and Cui, Zhu-Fang and Chang, Lei and Morgado, Jose-Manuel and Roberts, Craig D. and Rodriguez-Quintero, Jose",
    title = "{Revealing pion and kaon structure via generalised parton distributions *}",
    eprint = "2109.11686",
    archivePrefix = "arXiv",
    primaryClass = "hep-ph",
    reportNumber = "NJU-INP 051/21",
    doi = "10.1088/1674-1137/ac3071",
    journal = "Chin. Phys. C",
    volume = "46",
    number = "1",
    pages = "013105",
    year = "2022"
}

@article{Binosi:2016nme,
    author = "Binosi, Daniele and Mezrag, Cedric and Papavassiliou, Joannis and Roberts, Craig D. and Rodriguez-Quintero, Jose",
    title = "{Process-independent strong running coupling}",
    eprint = "1612.04835",
    archivePrefix = "arXiv",
    primaryClass = "nucl-th",
    doi = "10.1103/PhysRevD.96.054026",
    journal = "Phys. Rev. D",
    volume = "96",
    number = "5",
    pages = "054026",
    year = "2017"
}

@article{Qin:2011xq,
    author = "Qin, Si-xue and Chang, Lei and Liu, Yu-xin and Roberts, Craig D. and Wilson, David J.",
    title = "{Investigation of rainbow-ladder truncation for excited and exotic mesons}",
    eprint = "1109.3459",
    archivePrefix = "arXiv",
    primaryClass = "nucl-th",
    doi = "10.1103/PhysRevC.85.035202",
    journal = "Phys. Rev. C",
    volume = "85",
    pages = "035202",
    year = "2012"
}

@article{Bender:1996bb,
    author = "Bender, A. and Roberts, Craig D. and Von Smekal, L.",
    title = "{Goldstone theorem and diquark confinement beyond rainbow ladder approximation}",
    eprint = "nucl-th/9602012",
    archivePrefix = "arXiv",
    reportNumber = "ANL-PHY-8302-TH-96",
    doi = "10.1016/0370-2693(96)00372-3",
    journal = "Phys. Lett. B",
    volume = "380",
    pages = "7--12",
    year = "1996"
}

@article{Ferreira:2025anh,
    author = "Ferreira, M. N. and Papavassiliou, J.",
    title = "{Gluon mass scale through the Schwinger mechanism}",
    eprint = "2501.01080",
    archivePrefix = "arXiv",
    primaryClass = "hep-ph",
    doi = "10.1016/j.ppnp.2025.104186",
    journal = "Prog. Part. Nucl. Phys.",
    volume = "144",
    pages = "104186",
    year = "2025"
}

@article{Ding:2022ows,
    author = "Ding, Minghui and Roberts, Craig D. and Schmidt, Sebastian M.",
    title = "{Emergence of Hadron Mass and Structure}",
    eprint = "2211.07763",
    archivePrefix = "arXiv",
    primaryClass = "hep-ph",
    reportNumber = "NJU-INP 066/22",
    doi = "10.3390/particles6010004",
    journal = "Particles",
    volume = "6",
    number = "1",
    pages = "57--120",
    year = "2023"
}

@article{Schlessinger:1968vsk,
    author = "Schlessinger, L.",
    title = "{Use of Analyticity in the Calculation of Nonrelativistic Scattering Amplitudes}",
    doi = "10.1103/PhysRev.167.1411",
    journal = "Phys. Rev.",
    volume = "167",
    number = "5",
    pages = "1411",
    year = "1968"
}

@article{Schlessinger:1966zz,
    author = "Schlessinger, L. and Schwartz, C.",
    title = "{Analyticity as a Useful Computation Tool}",
    doi = "10.1103/PhysRevLett.16.1173",
    journal = "Phys. Rev. Lett.",
    volume = "16",
    pages = "1173--1174",
    year = "1966"
}

@article{Xu:2023bwv,
    author = "Xu, Yin-Zhen and Raya, Kh{\'e}pani and Cui, Zhu-Fang and Roberts, Craig D. and Rodr{\'\i}guez-Quintero, J.",
    title = "{Empirical Determination of the Pion Mass Distribution}",
    eprint = "2302.07361",
    archivePrefix = "arXiv",
    primaryClass = "hep-ph",
    reportNumber = "NJU-INP 070/23",
    doi = "10.1088/0256-307X/40/4/041201",
    journal = "Chin. Phys. Lett.",
    volume = "40",
    number = "4",
    pages = "041201",
    year = "2023"
}

@article{Bollweg:2025iol,
    author = "Bollweg, Dennis and Gao, Xiang and He, Jinchen and Mukherjee, Swagato and Zhao, Yong",
    title = "{Transverse-momentum-dependent pion structures from lattice QCD: Collins-Soper kernel, soft factor, TMDWF, and TMDPDF}",
    eprint = "2504.04625",
    archivePrefix = "arXiv",
    primaryClass = "hep-lat",
    doi = "10.1103/j3n6-8kxy",
    journal = "Phys. Rev. D",
    volume = "112",
    number = "3",
    pages = "034501",
    year = "2025"
}

@article{Lorce:2016ugb,
    author = "Lorc{\'e}, C. and Pasquini, B. and Schweitzer, P.",
    title = "{Transverse pion structure beyond leading twist in constituent models}",
    eprint = "1605.00815",
    archivePrefix = "arXiv",
    primaryClass = "hep-ph",
    doi = "10.1140/epjc/s10052-016-4257-8",
    journal = "Eur. Phys. J. C",
    volume = "76",
    number = "7",
    pages = "415",
    year = "2016"
}

@article{Anselmino:2005nn,
    author = "Anselmino, M. and Boglione, M. and D'Alesio, U. and Kotzinian, A. and Murgia, F. and Prokudin, A.",
    title = "{The Role of Cahn and sivers effects in deep inelastic scattering}",
    eprint = "hep-ph/0501196",
    archivePrefix = "arXiv",
    doi = "10.1103/PhysRevD.71.074006",
    journal = "Phys. Rev. D",
    volume = "71",
    pages = "074006",
    year = "2005"
}

@article{Collins:2005ie,
    author = "Collins, J. C. and Efremov, A. V. and Goeke, K. and Menzel, S. and Metz, A. and Schweitzer, P.",
    title = "{Sivers effect in semi-inclusive deeply inelastic scattering}",
    eprint = "hep-ph/0509076",
    archivePrefix = "arXiv",
    doi = "10.1103/PhysRevD.73.014021",
    journal = "Phys. Rev. D",
    volume = "73",
    pages = "014021",
    year = "2006"
}

@article{Raya:2024glv,
    author = "Raya, Kh{\'e}pani and Bashir, Adnan and Rodr{\'\i}guez-Quintero, Jos{\'e}",
    title = "{Mapping Spatial Distributions within Pseudoscalar Mesons}",
    eprint = "2412.06025",
    archivePrefix = "arXiv",
    primaryClass = "hep-ph",
    doi = "10.1088/0256-307X/42/2/020201",
    journal = "Chin. Phys. Lett.",
    volume = "42",
    number = "2",
    pages = "020201",
    year = "2025"
}

@article{Chang:2020kjj,
    author = "Chang, Lei and Raya, Kh{\'e}pani and Wang, Xiaobin",
    title = "{Pion Parton Distribution Function in Light-Front Holographic QCD}",
    eprint = "2001.07352",
    archivePrefix = "arXiv",
    primaryClass = "hep-ph",
    doi = "10.1088/1674-1137/abae52",
    journal = "Chin. Phys. C",
    volume = "44",
    number = "11",
    pages = "114105",
    year = "2020"
}

@article{Wang:2025usl,
    author = "Wang, Xiaobin and Chang, Lei and Ding, Minghui and Raya, Khepani and Roberts, Craig D.",
    title = "{Symmetry Constraints on Pion Valence Structure}",
    eprint = "2510.23950",
    archivePrefix = "arXiv",
    primaryClass = "hep-ph",
    reportNumber = "NJU-INP 106-25",
    month = "10",
    year = "2025"
}

\end{document}